\documentclass[10pt,journal,compsoc]{IEEEtran}

\IEEEoverridecommandlockouts                              
 \pdfoutput=1   
\usepackage{graphicx}     
\usepackage{epsfig} 
\usepackage{epstopdf}
\usepackage{amssymb}
\usepackage{amsmath} 
\usepackage{pgf}
\usepackage{caption}
\usepackage[caption=false]{subfig}
\usepackage[export]{adjustbox}
\usepackage{algorithm}
\usepackage{algorithmicx,algpseudocode}
\usepackage[normalem]{ulem}
\usepackage{lipsum}
\usepackage{mathtools}
\usepackage{cuted}
\usepackage{multirow}
\usepackage{float}
\usepackage{cite}
\usepackage{booktabs}
\usepackage{multirow}
\usepackage{color}
\usepackage{bm}
\newcommand{\comment}[1]{}

\newcommand\dan[1]{\textcolor{black}{#1}}
\begin{document}

\title{\LARGE \bf
Distributed Deep Reinforcement Learning for Functional Split Control in Energy Harvesting Virtualized Small Cells
}

\author{
    \IEEEauthorblockN{Dagnachew Azene Temesgene\IEEEauthorrefmark{1}, Marco Miozzo\IEEEauthorrefmark{1}, Deniz G{\"u}nd{\"u}z \IEEEauthorrefmark{2}, Paolo Dini\IEEEauthorrefmark{1}}\\
    \IEEEauthorblockA{\IEEEauthorrefmark{1}CTTC/CERCA\\ Av. Carl Friedrich Gauss, 7, 08860, Castelldefels, Barcelona, Spain
    \\\{dagnachew.temesgene,marco.miozzo, paolo.dini\}@cttc.es}\\
    \IEEEauthorblockA{\IEEEauthorrefmark{2}Imperial College London, London, SW7 2AZ, UK
    \\\ d.gunduz@imperial.ac.uk}
}
\IEEEtitleabstractindextext{
\begin{abstract}
To meet the growing quest for enhanced network capacity, mobile network operators (MNOs) are deploying dense infrastructures of small cells. This, in turn, increases the power consumption of mobile networks, thus impacting the environment. As a result, we have seen a recent trend of powering mobile networks with harvested ambient energy to achieve both environmental and cost benefits.  In this paper, we consider a network of virtualized small cells (vSCs) powered by energy harvesters and equipped with rechargeable batteries, which can opportunistically offload baseband (BB) functions to a grid-connected edge server depending on their energy availability. We formulate the corresponding grid energy and traffic drop rate minimization problem, and propose a distributed deep reinforcement learning (DDRL) solution. Coordination among vSCs is enabled via the exchange of battery state information. The evaluation of the network performance in terms of grid energy consumption and traffic drop rate confirms that enabling coordination among the vSCs via knowledge exchange achieves a performance  close to the optimal. Numerical results also confirm that the proposed DDRL solution provides higher network performance, better adaptation to the changing environment, and higher cost savings with respect to a tabular multi-agent reinforcement learning (MRL) solution used as a benchmark. 
\end{abstract}

}
\maketitle

\begin{table}[H]
  \centering
  \caption{List of frequently used acronyms and symbols}
  \begin{tabular}{l c}
   \hline 
   BB & baseband\\
   BS & base station \\
   DDRL & distributed deep reinforcement learning\\
   DP & dynamic programming\\
   EH &  energy harvesting\\
   FQL & fuzzy Q-learning\\
   MBS & macro base stations\\
   MDP & Markov decision process \\
   MEC & multi-access edge computing \\
   MRL & multi-agent reinforcement learning \\
  NFV & network function virtualization\\
   RAN & radio access network\\
   RL & reinforcement learning \\
   SBS & small base station\\ 
   SDN & software defined networking\\
   SGD & stochastic gradient descent\\
   vSC & virtualized small cell\\
   $\boldsymbol{A}^t$ & Operative states (control actions) of the vSCs in slot $t$ \\
   $\boldsymbol{B}^t$ & Energy stored in batteries at beginning of slot $t$ \\
   $\boldsymbol{H}^t$ & Energy harvested by vSCs in slot $t$\\
   ${h}^t$ & Hour of the day in slot $t$\\
   ${m}^t$ & Month in slot $t$\\
   $\boldsymbol{L}^t$ & Traffic load generated inside coverage of vSCs in slot $t$\\
   $r_t$ & Scalar reward signal \\
   $\boldsymbol{X}^t$ & State of the vSCs in slot $t$ \\
  $\alpha$ & Learning rate \\
  $\varepsilon$ & Exploration parameter \\
  $\gamma$ & Discount factor \\
  \hline
  \end{tabular}
\label{tab:acronym_table}
\end{table}
\section{Introduction}
Due to an exponential growth in mobile traffic demand \cite{cisco}, dense heterogeneous networks (HetNets) of multi-tier base stations (BSs) are being deployed as a means of enhancing capacity. In a HetNet, a large number of small BSs (SBSs) are employed to ensure coverage of hot-spots (e.g., entertainment areas, shopping malls, offices), while macro BSs (MBSs) are deployed for ensuring mobility and coverage.  One of the issues that arise from the dense deployment of SBSs is the rapidly increasing electrical power consumption, which is playing a major part in the operational expenditures of mobile network operators (MNOs) \cite{footprint}. Hence, energy sustainable design and operation of mobile networks is identified as one of the key requirements of 5G and beyond mobile networks in order to ensure cost effectiveness and reduce the impact on the environment. To this end, new architectural paradigms, such as multi-access edge computing (MEC), are emerging, and renewable energy is gaining popularity as a means to decrease dependency on the power grid \cite{ehcommag}.


MEC enables BSs to leverage cloud computing capabilities and offers computational resources on-demand basis \cite{mec}. Relying on MEC, a \textit{flexible functional split} between SBSs and a centralized baseband (BB) unit pool~\cite{scforum} has been proposed, where part of the BB processes are executed at the SBSs, while the remainder is offloaded to a central BB unit (BBU) pool. This solution can be enabled via network function virtualization (NFV), which permits network functions to be executed on general purpose hardware as virtual functions, and software defined networking (SDN) as a tool to manage these functions~\cite{sdn-nfv}. As a result, network functions of SBSs can be virtualized and placed at different sites of the network. These SBSs, known as virtualized small cells (vSCs), enable higher flexibility in resource allocation and management. 

\dan{In parallel, energy harvesting (EH) technology is becoming widely applicable in mobile networks. EH allows both cost and environmental impact reduction~\cite{piro2013}; however, it comes with its own unique challenges, mainly due to unreliable intermittent energy sources. Hence, in EH BSs, it is important to intelligently manage the harvested energy in order to avoid service degradation or interruption. Consequently, BSs with EH and MEC can open a new frontier in energy-aware processing and sharing of resources according to flexible functional splits. 
The EH powered vSCs can opportunistically use the processing capabilities of the MEC server, which can be co-located at the MBS site. In particular, we consider a two-tier scenario where EH powered vSCs can offload part of their BB processing to the central MBS site with a MEC server. This is particularly important since the power consumption due to BB processing has a huge share in the total power consumption breakdown of SBSs~\cite{earth-D23}.}
In \cite{fss-vtc}, we have proposed an offline solution for performance bounds of dynamic selection of functional split options for vSCs powered by EH. These results prove that dynamically adapting functional split options can provide significant grid energy savings as opposed to static configurations. However, the offline solution adopted relies on \textit{a-priori} knowledge and does not scale up with the number of vSCs due to high computational complexity. 

\dan{On the other hand, reinforcement learning (RL) allows learning an optimal/near optimal strategy through interactions with the environment while achieving a system wide goal, e.g., efficient utilization of the harvested energy, without requiring the model of the environment variables (e.g., user demands, EH).} However, implementing RL algorithms in the presence of multiple vSCs operating in parallel is challenging. Centralized solutions experience long convergence and training phases due to prohibitively large state/action sets. A distributed approach may allow to reduce the complexity by dividing the problem among multiple agents. A common reward signal is employed so that the agents aim at optimizing the same system wide goal in a distributed manner.
This approach, also known as multi-agent RL (MRL)~\cite{marl-survey}, scales better due to the distribution of the learning and decision making processes among the vSCs. However, in MRL, the impact of each individual agent's actions on the reward is difficult to distinguish as the reward may depend on all the actions in a complex manner. 
Hence, MRL solutions should ensure some coordination among the agents (i.e., the vSCs) towards achieving system wide gains.

\dan{MRL based algorithms for dynamic selection of functional split options in vSCs with EH capabilities are proposed in our previous work \cite{fss-comcom}. In \cite{fss-comcom}, distributed Q-learning (QL) and fuzzy Q-learning (FQL) are applied with coordination enabled via broadcasting the normalized traffic load of the MBS. In this work, we propose to achieve better coordination among the agents through the exchange of specific local state information.} Hence, we increase the amount of shared knowledge in order to reduce conflicting behaviors, and, in turn, converge to stationary policies with higher system wide gains. The main challenge with this approach is the exponential increase in the state space dimension, which may slow down the learning process and even jeopardize its convergence. Tabular MRL methods (e.g., QL, FQL) work through mapping each state to a value; hence, each state-action pair needs to be properly explored. \dan{In problems with continuous state variables, as ours, this mapping is performed through quantization or fuzzy inference systems, and may result in large state spaces.  For instance, the solutions proposed in \cite{fss-comcom} rely on broadcasting the MBS traffic load and have $4$ state variables corresponding to energy, battery, local and MBS traffic load. Denoting the quantization level by $z$, the state space of each agent in \cite{fss-comcom} has size $z^4$. If vSCs exchange battery state information for better coordination, the number of states will be multiplied by a factor of $z^N$, where $N$ is the number of vSCs in the system. This implies an increment in the size of the state space from $625$ in \cite{fss-comcom} to a range from $78125$ to $1.9073486 \times e^{13}$ for $z=5$, in a scenario of $3$ and $15$ vSCs, respectively}. Tabular MRL methods cannot be applied on such a large state space.

In deep RL (DRL), deep neural networks are used to approximate the Q-values of state-action pairs \cite{atari}. DRL allows working with large state and action spaces through Q-value estimations without the need for large and impractical look-up tables. Accordingly, \dan{we propose a distributed DRL (DDRL) algorithm for the dynamic control of functional split options in vSCs with EH capabilities, where each vSC is modeled as a distinct DRL-based agent that takes decisions in coordination with other vSC agents.} As opposed to tabular MRL, DDRL allows to coordinate the policies of the learning agents via local state information exchanges without facing practically infeasible state-action tables.

\dan{The main contributions of the paper are summarized as follows: }
\begin{itemize}
	                  
	\item \dan{We formulate a network wide sequential decision making problem in order to optimally leverage flexible functional split options at the vSCs with the goal of minimizing both the grid energy consumption and the amount of dropped traffic. }

	\item \dan{We propose a MRL solution, i.e., DDRL, that can handle the prohibitively large state space in an efficient manner. We analyze its complexity and convergence properties and describe its spatio-temporal behavior.}
        
	\item \dan{We evaluate the performance (in terms of energy consumption and traffic drop rate) of the proposed DDRL solution and compare it against benchmark multi-agent FQL \cite{fss-comcom} as well as an offline performance bound \cite{fss-vtc}. In addition, energy and cost savings of MRL-based controllers are estimated as compared to a system relying only on grid power.}
     
\end{itemize}

The rest of the paper is organized as follows. Section~\ref{sec:relatedwork} describes the related literature. Section~\ref{sec:architecture} presents the reference architecture considered in this work. Section~\ref{sec:system-model} describes the the problem statement as well as power consumption, traffic and EH models. The proposed DDRL solution is explained in Section~\ref{sec:control}.  Section~\ref{sec:results} is dedicated to the simulation scenario, numerical results of simulations including the comparison with an FQL solution and cost analysis. Finally, we draw our conclusions in Section~\ref{sec:conclusions}.

\section{Related work}
\label{sec:relatedwork}

As a result of dense deployment of BSs combined with the increasing importance of energy sustainability, intelligent energy management in EH BSs has been the focus of many recent studies.  Most of this literature analyzes hierarchical multi-tier networks, the so called HetNets, with an intelligent switching on/off scheduling of BSs. The authors in~\cite{piovesan2017} apply dynamic programming (DP) to determine the optimal switch on/off policy in a two-tier HetNet with baseline MBS and hot-spot deployed SBSs. The solution shows the performance bound of an intelligent switch on/off policy when all the system dynamics information are known a-priori. Minimizing the grid energy consumption for hybrid powered BSs is also studied in \cite{gong2014}. Here, the authors apply two-stage DP designed to achieve energy saving gains while maintaining the probability of blocking. The authors in \cite{zhou2013} study sleep mode coordination between BSs powered by EH and grid energy using DP. However, the DP-based solution is shown to entail high computational complexity. 

The authors in~\cite{lee2017} apply a ski-rental framework based on-line algorithm for optimal switch on/off scheduling for minimizing the operational costs of a network composed of self-powered BSs. The application of QL for the optimization of an EH system is studied in \cite{ehql}. The authors in~\cite{miozzo2015} apply distributed QL in the context of HetNets to optimize the harvested energy utilization. Multi-armed bandit based distributed learning is applied in \cite{ehmab, ameur2016} to allow each SBS to learn its own energy-efficient policy. The authors in \cite{ll-vtc, LL} apply layered learning for system wide harvested energy allocation through decomposition of the problem into two layers. The first layer, based on RL, is in charge of local control at each SC, while the second layer, based on neural networks, ensures network wide coordination among the SCs. The authors in \cite{drag} applied deep RL methods for the minimization of energy consumption in HetNets through optimal activation of a subset of the SCs, while maintaining the desired level of QoS. In particular, they have applied actor-critic RL methods where deep neural networks are used as policy and value function approximators. The authors in \cite{mendil} proposed a RL-based energy controller for a SC powered by EH, battery and smart grid by considering battery ageing effects. This work is based on FQL and is shown to provide significant extension to the life time of a small cell battery. Renewable energy allocation in edge computing devices with EH is studied in \cite{xu2016}. Here, the authors propose RL-based online solutions for offloading and auto-scaling in edge computing devices that are powered by EH. RL based algorithms for dynamic placement of functional split options is proposed in \cite{dynamic-pimrc}. It is based on temporal difference (TD) learning methods, namely QL and SARSA for online learning of control policies of a vSC powered by EH with flexible operative modes.  On the other hand, the authors in \cite{wangflexible} propose an optimal flexible functional split option selection scheme for a cloud RAN with radio remote units supplied by renewable energy sources. They show that the optimal functional split selection problem can be formulated as convex optimization, and propose a heuristic online algorithm without relying on future energy arrival information. However, the study in \cite{wangflexible} focuses only on throughput maximization for a single remote radio unit. Indeed, most of the literature on EH with MEC focus on a single SC scenario, and the literature on multiple SCs consider only on/off switching policies. Here, we study more configuration options of SCs, in addition to switch on/off and enable higher grid energy savings. 

In \cite{fss-comcom}, we propose a MRL solution for the dynamic control of functional split options ensuring scalable solutions. In particular, it is based on distributed tabular RL algorithms, i.e., QL and FQL.  This paper extends the work in \cite{fss-comcom} by proposing a multi-agent DRL solution that overcomes the limitations of tabular MRL approaches. We propose coordinated control via communication of battery level information among multiple learning agents and a common system wide reward signal. Moreover, we tailor distributed DRL algorithm to our network scenario and evaluate its performance against a benchmark multi-agent FQL controller studied in \cite{fss-comcom}.
\begin{figure}[b!]
	\centering
	\includegraphics[scale=1.6]{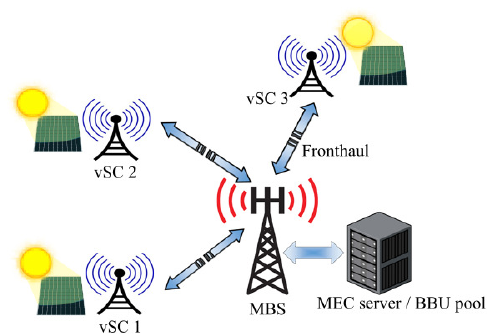}
	\caption{Reference two-tier network architecture.}
	\label{fig:scenario}
\end{figure}
\section{Reference Architecture}
\label{sec:architecture}

This work considers a two-tier network architecture illustrated in Figure \ref{fig:scenario}. The first tier consists of a MBS and a co-located BBU pool, acting as the MEC server. MBS is responsible for providing baseline coverage, mobility support, and BB processing resources. The MBS site is fully powered by the grid, thus assuring reliable communications and computing. The second tier is composed of vSCs, which do not overlap in coverage \cite{ngmn}. They are powered solely by solar panels and are equipped with finite-capacity batteries.  

\dan{The vSCs opportunistically employ central BBU pool for full or partial BB processing according to flexible functional splits. 
3GPP has defined different functional splits between the distributed and centralized units~\cite{3gpp-38801}, which correspond to vSCs and the MBS, respectively.}
In our model, vSCs can opportunistically operate in one of the following functional split configurations specified in~\cite{3gpp-38801}: 
\begin{itemize}
	\item PHY-RF split: all the protocols, physical (PHY) and higher layers, are implemented at the MEC server. Hence, the vSC behaves as a radio frequency (RF) transceiver, used only for signal transmission and reception;
	\item MAC-PHY split: PHY layer processing takes place at the vSC, in addition to RF functions.  Medium access control (MAC) and higher layer functions are executed at the MEC server. 
\end{itemize}

\begin{figure}[t!]
	\centering
	\includegraphics[scale=0.5]{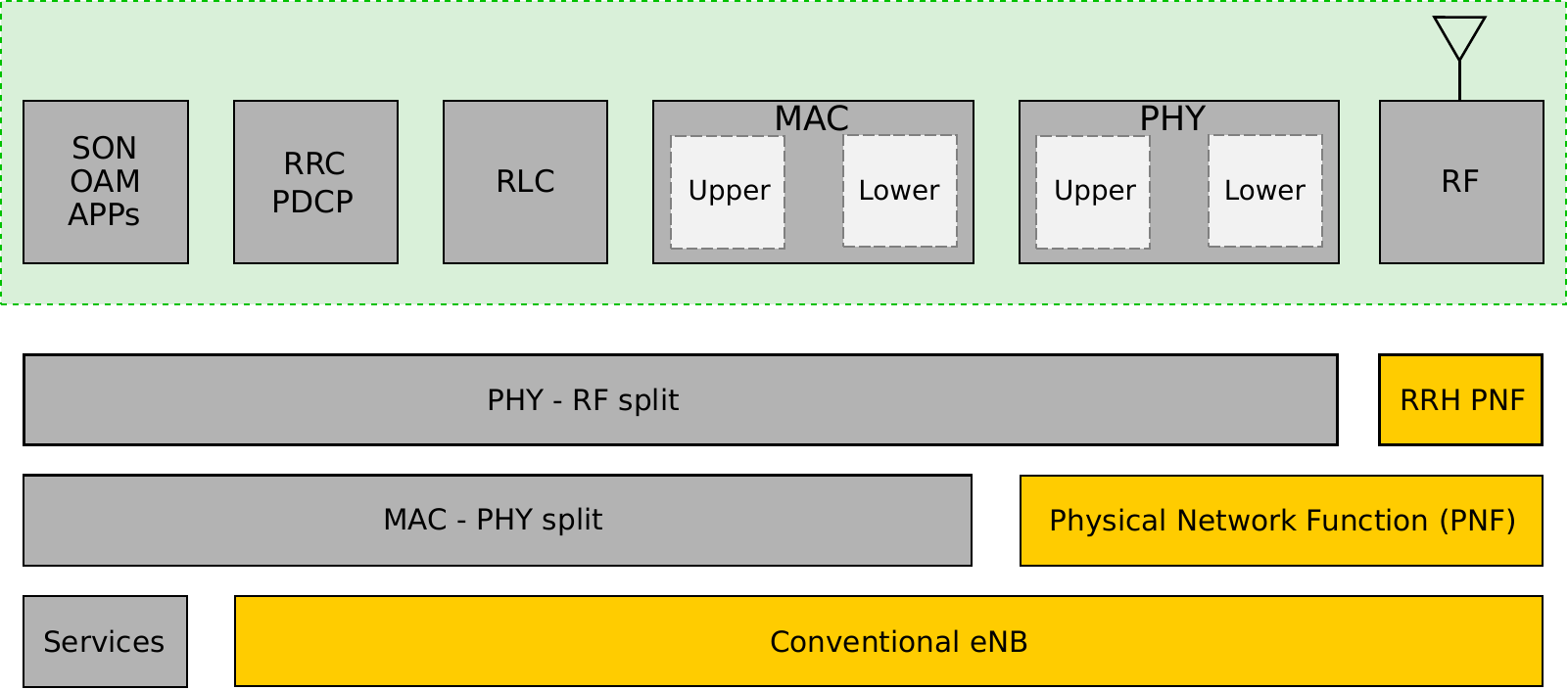}
	\caption{Different implementations of the functional split configurations including PHY-RF and MAC-PHY split. The conventional eNodeB (eNB) configuration is also shown for comparison.}
	\label{fig:split}
\end{figure}

These two functional split options have been selected based on their impact on the energy consumption of vSCs.  PHY-RF and MAC-PHY split options impose significantly different energy requirements on the vSCs, which allows to implement a dynamic control on local energy consumption. Other functional split options have negligible impact on the energy utilization, as demonstrated in ~\cite{fss-vtc}.
These functional split options are depicted in Figure~\ref{fig:split} along with the conventional eNodeB architecture. Each option corresponds to a different computational load for the vSCs and MBSs, which in turn, corresponds to different energy consumption models, as will be described in Section~\ref{sec:bs-power-model}. 
\section{ System Model}
\label{sec:system-model}

\subsection{Problem statement}
\label{sec:network-model}
We consider a two-tier mobile network composed of one MBS with a co-located BBU pool and $N$ vSCs.
The network is modeled as a discrete time dynamic system, which evolves in time slots based on the variations in the traffic demand and energy arrivals. The traffic loads at time slot $t$ generated by the users in the coverage area of the vSCs are denoted by $\boldsymbol{L}^t \triangleq [L_1^t,L_2^t,\ldots,L_N^t ]$, where $L_n^t$ is the traffic load at the $n\textsuperscript{th}$ vSC.  The energy harvested by the vSCs in slot $t$ are denoted by $\boldsymbol{H}^t \triangleq [H_1^t,H_2^t,\ldots,H_N^t ]$, while the battery states are denoted by $\boldsymbol{B}^t \triangleq [B_1^t,B_2^t,\ldots,B_N^t ]$, where $H_n^t$ and $B_n^t$ are the harvested energy and the battery state of the $n\textsuperscript{th}$ vSC in time slot $t$, respectively. In addition, in order to capture the evolution of the traffic requests and energy arrivals, the hour of the day and the month are defined as ${h}^t$ and ${m}^t$, respectively, for time slot $t$. 
Battery states evolve according to the following relation:
\begin{equation}
\boldsymbol{B}^{t+1}=\min \left(\boldsymbol{B}^t+\boldsymbol{H}^t-\boldsymbol{P}^t\Delta_t,B_{\mathrm{cap}}\right)
\label{eq:battery}
\end{equation}
where $\boldsymbol{P}^t \triangleq [P_1^t,P_2^t,\ldots,P_N^t ]$ and $P_n^t$  is the power consumed by the $n\textsuperscript{th}$ vSCs in slot $t$ (will be described in detail in Section~\ref{sec:bs-power-model}), $B_{\mathrm{cap}}$ is the battery capacity, and  $\Delta_t$ is the duration of one time slot.
At each time slot, each vSC can be in one of the three modes of operation. Denoting the mode of vSCs by $\boldsymbol{A}^t$, the mode of $n\textsuperscript{th}$ vSC in time slot $t$, $A_n^t$, is given by:
\begin{equation}
A_n^t =
\begin{cases}
0 \quad \textrm{if the } n \textrm{-th vSC is OFF}\\
1 \quad \textrm{if the } n \textrm{-th vSC is in PHY-RF mode}\\
2 \quad \textrm{if the } n \textrm{-th vSC is in MAC-PHY mode}\\
\end{cases}
\end{equation}

The network wide sequential decision making problem is defined by a Markov decision process (MDP) as $\boldsymbol{X}^{t+1}=f(\boldsymbol{X}^t,\boldsymbol{A}^t,\boldsymbol{L}^t, \boldsymbol{H}^t)$, where $\boldsymbol{X}^t \triangleq [X_1^t,X_2^t,\ldots,X_N^t ]$ denotes the states of the vSCs in slot $t$, $\boldsymbol{A}^t \triangleq [A_1^t,A_2^t,\ldots,A_N^t ]$ are the control actions/modes of the vSCs, and ($\boldsymbol{L}^t, \boldsymbol{H}^t$) are the environmental random variables (i.e., traffic and EH stochastic processes). In particular, we define each state $X_i^t$, $i=1,...,N$, as $X_i^t = (h^t, m^t,  L^t_i, \boldsymbol{B^t})$. Hence, the state of the $i\textsuperscript{th}$ vSC in slot $t$ is represented by the battery levels of each vSC $\boldsymbol{B^t}$, its traffic load  $L^t_i$, the month of operation $m^t$ and the hour of the day $h^t$.

 
We have two objectives, i.e., to minimize the grid energy consumption at the MBS, and to minimize the traffic demands that cannot be satisfied due to vSCs being in the OFF mode. Hence, the optimization goal at every decision slot $t$ will be to minimize the total weighted cost over a finite time horizon $T$, given by:
\begin{equation}
\begin{split}
\text{\boldmath{P1:}}\min_{\{\boldsymbol{A}^t\}_{t=i,\dots,T+i}}\; &\sum_{t=i}^{T+i} \omega\cdot \mathrm{E}(\boldsymbol{A}^t)+ (1 - \omega)\cdot {\mathrm{D}}(\boldsymbol{A}^t),\\
\\
\end{split}
\label{eq:optimization}
\end{equation}
where $t=i$ refers to the $i\textsuperscript{th}$ decision slot, $\mathrm{E_m}(\boldsymbol{A}^t)$ and  $\mathrm{D}(\boldsymbol{A}^t)$ denote, respectively, the normalized grid energy consumption and the traffic drop rates, which are defined next. The grid energy consumption is equivalent to the energy consumption at the MBS site as the vSCs depend solely on harvested energy. The grid energy consumption is normalized with respect to the maximum possible energy consumed by the MBS, i.e., when it is at full load and performing the BB processes of vSCs in PHY-RF mode. 
Hence, the normalized grid energy consumption is computed as:  
\begin{equation}
\mathrm{E}(\boldsymbol{A}^t)= \frac{\mathrm{P}(\boldsymbol{A}^t)}{\mathrm{P^{MAX}}},
\label{eq:mbs-energy}
\end{equation}
where $\mathrm{P}(\boldsymbol{A}^t)$ is the power consumption of the MBS given the operative modes of the vSCs at slot $t$, and $\mathrm{P}^\mathrm{MAX}$ is the power consumption of the MBS at full load.
The details of the power consumption model are described in Section \ref{sec:bs-power-model}.
The traffic drop rate, $\mathrm{D}(\boldsymbol{A}^t)$, is the ratio of the total traffic demand that cannot be served in slot $t$. Additionally, we impose the state of charge (SOC) of the batteries to be maintained above a threshold $B_{\mathrm{th}}$ to avoid a rapid reduction in lifetime~\cite{Lu2013}.
The weight $\omega$ determines the balance between the two objectives. In this work, we consider $\omega = 0.5$ to impose equal importance on the two objectives, but the results can easily be generalized to arbitrary weights.

\dan{A centralized offline solution is proposed in~\cite{fss-vtc} using DP and with \textit{a priori} knowledge of the system, i.e., assuming that all future energy and traffic arrivals are known. 
Those results can be considered as performance bounds and used as a benchmark for the policies proposed in this paper.
In this work, we instead propose an on-line solution based on DDRL, without assuming the explicit knowledge of the system statistics governing the underlying random processes. In particular, we propose DDRL agents in which neural networks are used as value approximation functions \cite{atari} to determine the optimal actions $\boldsymbol{A}^t$.  Our proposal is based on distributed and coordinated decision making, i.e., each vSC takes its own action based on its state, which makes it scalable with the number of vSCs. In order to coordinate the decision making process, we rely on partial local state information exchange among vSCs; in particular, we assume that each vSC knows the battery levels of all the other vSCs. 
Section \ref{sec:control} describes the proposed DDRL solution to the sequential decision making process described here.}

\subsection{Power model}
\label{sec:bs-power-model}
The power consumption of each split option is estimated based on the model introduced in \cite{desset2012}, which is a general flexible power model of BSs and provides the power consumption in giga operations per second (GOPS). Technology-dependent GOPS to Watt conversion factor is applied to determine the power consumption in Watts. 
In this paper, we have mapped the various BB processing tasks of the functional split options to their power requirement estimations.

The total BS power consumption is given by:
\begin{equation}
P_{\mathrm{BS}}=P_{\mathrm{BB}}+P_{\mathrm{RF}}+P_{\mathrm{PA}} + P_{\mathrm{overhead}},
\label{eq:Ptotal}
\end{equation}
where $P_{\mathrm{BB}}$ is the power consumption due to BB processing, $P_{\mathrm{RF}}$ is the power consumption of RF circuitry, $P_{\mathrm{PA}}$ is the power consumption by the power amplifier, and $P_{\mathrm{overhead}}$ is the overhead power consumption (e.g., cooling system). BB power consumption, $P_{BB}$, consists of the idle mode power consumption and the power consumption OFDM processing, filtering, frequency domain processing, and forward error correction. In accordance with~\cite{desset2012}, all these components scale with the number of antennas and bandwidth and the power consumption of frequency domain processing and forward error correction also depend on the traffic load. When the vSCs are in PHY-RF split mode, their power consumption model does not include the corresponding $P_{BB}$ which is added to the power consumption of the MBS,  since the BB processing takes place at the MBS site. On the other hand, in MAC-PHY split mode, the vSC power consumption includes the  $P_{BB}$  term, and is given by (\ref{eq:Ptotal}).
 Considering the aforementioned model description, the grid power consumed by the MBS is computed as:

\begin{equation}
P_\mathrm{m} =  P_\mathrm{BS}^\mathrm{MBS} + \sum_{i \in \mathcal{G}}P_\mathrm{BB}^i,
\label{eq:PBBsbs}
\end{equation}
where $P_\mathrm{BS}^\mathrm{MBS}$ is the power consumption of the MBS computed as in (\ref{eq:Ptotal}), $P_\mathrm{BB}^i$ is the BB power consumption of the $i$-th vSC, and $\mathcal{G}$ is the set containing the indexes of the vSCs in PHY-RF split mode. 

\begin{figure*}[ht]
\centering
\hspace*{-0.0cm}\includegraphics[scale = 0.6]{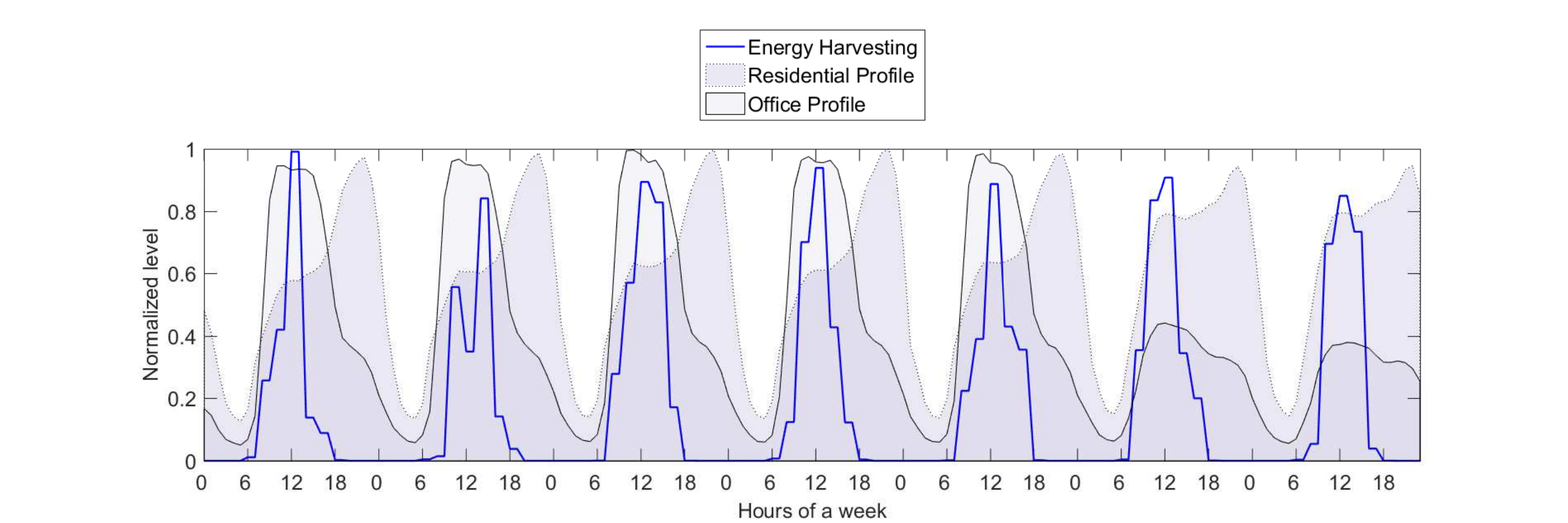}
\caption{Typical weekly normalized EH, office traffic and residential traffic profiles. } 
\label{fig:traffic-eh}
\end{figure*}

\subsection{EH and Demand Profiles}
\label{sec:eh-traffic-profiles}
EH and traffic demand are typically modeled as time-correlated random processes. Many works have focused on obtaining accurate models of these random processes. 
Instead, in this work, we directly use hourly energy generation traces from a solar source for the city of Los Angeles (CA, USA). The solar raw irradiance data has been collected from the national renewable energy laboratory and converted into harvested energy traces using the SolarStat tool~\cite{Miozzo2014}. EH traces are generally bell-shaped with a peak around midday, whereas the energy harvested during night is negligible. Moreover, as discussed in~\cite{Miozzo2014}, high variability of the harvested energy may occur during the day, even in summer months. As a result, although the energy inflow pattern can be known to a certain extent, intelligent and adaptive algorithms that make their decisions based on current and past inflow patterns, as well as predictions of future energy arrivals, have to be designed. 

For the demand profile, the UEs have been classified as \emph{heavy} and \emph{ordinary} users according to their amount of requested traffic ~\cite{earth-D23}. 
The traffic demand of each UE with in a time slot are estimated to resemble realistic traffic profiles presented in~\cite{Xu2017understanding}, which are derived from time, location and frequency information of thousands of cellular towers. The analysis in~\cite{Xu2017understanding} demonstrates that the urban mobile traffic usage can be described by mainly five basic time domain patterns that correspond to different functional regions, i.e., residential, office, transportation, entertainment, and comprehensive. In this article, we are considering residential and office profiles, which are the most common use cases for urban deployment scenarios. In addition, based on the average traffic generated by the users, traffic variability is added following a normal distribution using standard deviation from measurements of real mobile traffic traces~\cite{mobiletraffic}. Based on the energy traces~\cite{Miozzo2014} and traffic profiles~\cite{Xu2017understanding, earth-D23 }, an example of a normalized EH trace, residential traffic profile and office traffic profile for both week and weekend days is shown in Figure \ref{fig:traffic-eh}. The figure shows that EH and residential traffic profiles peak at different hours of the day, i.e., energy harvesting peak occurs around noon whereas traffic demand peak occurs during the evening. This calls for intelligent energy management to maximize the utilization of harvested energy.

\section{Distributed Deep Reinforcement Learning (DDRL)}
\label{sec:control}

\dan{In this section, we introduce the proposed DDRL algorithm where each vSC is modeled as a DRL agent taking decisions in coordination with other vSC agents. The section starts with background information on RL, followed by the details of the DDRL controllers including the states, actions and the reward function.}
\subsection{Background}
\label{sec:background}

RL control relies on learning by interacting with the environment without an exemplary supervision \cite{suttonRL}. It is a well known framework for solving problems described as MDPs. Formally, the RL framework is defined in terms of states, actions and rewards. Through the RL process, the agent executes a certain action according to its current state, and as a result of its action, it receives an immediate reward, and transitions to a new state. It is important to note that in RL, the rewards can be delayed. Hence, it is a sequential decision making process with the goal of maximizing cumulative reward. In our problem, the objective of the RL-based controller is to learn the energy management policy through interactions with the environment based on the traffic load, energy arrival, and battery state information. 
For a network with one vSC, let $\boldsymbol{X}^t$  denote the state at time $t$. The agent chooses an action $A^t$ from set $\boldsymbol{A^t}$, which is equivalent to the operative mode of the vSC. As a result of this action, the environment returns a reward $r_t$, which is used to update the corresponding Q-value, $Q(\boldsymbol{X}^t,A^t)$, which represents the estimate of the long term average discounted reward when taking a specific action at a given state. One of the most widely used RL algorithms is QL, which  is an off-policy method that can learn the optimal Q-values for each state-action pair \cite{suttonRL}. 
In QL, the Q-values of each state-action pair are updated as:
\begin{equation}
\begin{split}
&Q(\boldsymbol{X}^t,A^t) = Q(\boldsymbol{X}^t,A^t) + \\
&  \alpha (r_t + \gamma \max_{\boldsymbol{A}}Q(\boldsymbol{X}^{t+1},A) - Q(\boldsymbol{X}^t,A^t)),
\label{eq:q}
\end{split}
\end{equation}
where $\alpha$ is the learning rate, $\gamma$ is the discount factor, $A^t$ is the current action, $r_t$ is the immediate reward, $\boldsymbol{X}^t$ and $\boldsymbol{X}^{t+1}$ are the current and the next state, respectively. The process of learning needs to balance between \emph{exploration}, i.e., taking random actions to discover new knowledge, and \emph{exploitation}, i.e., taking an action that has been already discovered as good (an action with the highest Q-value). \emph{For the single-agent RL problem,} as long as all the state-action pairs are visited and updated infinitely often, QL is guaranteed to converge to an optimal policy regardless of the specific policy being followed throughout the learning phase. On the other hand,  in MRL, the learning and decision making processes are distributed among the agents. MRL is a more scalable approach suitable for complex systems, as it divides the problem among multiple agents. However, MRL has no formal proof of convergence to the optimal solution due to the non-stationarity arising from the simultaneous learning of the agents.

\subsection{DDRL-based control}
\label{sec:ddrl}
\dan{We follow a distributed design, where each vSC implements a DRL-based agent and acts independently but in coordination with the other vSC agents. The distributed design ensures scalability while maintaining the complexity of the controllers to a reasonable level by avoiding the exponentially growing number of actions that requires Q-value estimations. For instance, if there are $3$ possible operative modes/actions per vSC, there are $3^N$ possible combinations of operative modes, where $N$ is the number of vSCs, resulting in potentially different network performance, i.e., gird energy consumption and traffic drop rate. Hence, a centralized approach would require the exploration of all $3^N$ possible actions. This means that the number of Q-values that must be estimated reach $14,348,907$ for $15$ vSCs. In our DDRL implementation instead, each vSC takes its own actions and coordination among the agents is enabled by the exchange of battery state information as well as via a global reward signal, where each vSC receives the same system level feedback.}

\subsubsection{States}
\label{sec:states}
According to the system model defined in Section~\ref{sec:network-model}, the state of the \textit{$i^{th}$} vSC at time slot $t$ is defined as, $X_i^t = (h^t, m^t,  L^t_i, \boldsymbol{B^t})$, where $h^t$ denotes the hour of the day, $m^t$ denotes the month, $L_i^t$ denotes the traffic load experienced by the vSC and $\boldsymbol{B^t}$ denotes the vector of battery states of all vSCs. The values of input traffic load and the battery state variables, namely $ L_i^t$ and $\boldsymbol{B^t}$ are all normalized with respect to their maximum. On the other hand, for cyclic inputs $h^t$ and $m^t$, sinusoidal transformation is applied \cite{cyclic}. Hence, the hour values ranging from $[0,23]$ and months from $[0,11]$ are transformed into sinusoidal values between $[-1,1]$ and their cyclic properties are maintained. These state variables, after normalization, are the input of the neural network that is used to estimate the Q values of all the actions at that state. Since the battery levels of all the vSCs are part of the states, the size of the input to the neural network of each vSC agent is dependent on the number of vSCs and is given as $3 + N$, where $N$ is the number of vSC agents. 
\subsubsection{Actions}
The set of possible actions are the possible operative modes of the vSCs, $\boldsymbol{A}^t$. The action set for the \textit{$i^{th}$} vSC are switching off, PHY-RF split mode, or MAC-PHY split mode. Hence, the action set for the whole DDRL solution is a combination of the three operative modes of each vSC.
\subsubsection{Reward}
The reward function determines the immediate reward each DRL-agent acquires as a result of taking a specific action. The optimization goal is to minimize the power drained from the grid while reducing the system drop rate, as given by (\ref{eq:optimization}). Hence, the reward function can be formulated as:
\begin{equation}
r_t = 1 -(\omega \cdot \mathrm{E_m}(\boldsymbol{A}^t)+ (1 - \omega)\cdot {\mathrm{D}}(\boldsymbol{A}^t))
\label{eq:reward}
\end{equation}
where $\mathrm{E_m}(\boldsymbol{A}^t)$ were defined in Section \ref{sec:network-model}. In our DDRL implementation, each agent receives the same reward signal in (\ref{eq:reward}) to enable coordinated learning. 

In order to improve the convergence behavior of the DRL algorithm, we will use an experience buffer of capacity $M$, denoted by $D$, in which we store previous experiences of the agents consisting of the state, action, reward and next state tuples. We then randomly sample mini-batches from this experience buffer for training the neural networks of DRL agents. This technique is known as \textit{experience replay}. Neural networks used for estimating the Q-values are initialized with random weights, and an $\epsilon$-greedy policy maps the input states to actions; where random actions are chosen with probability of $\epsilon$, i.e., \textit{exploration}, otherwise an action with the highest Q-value is taken, i.e., \textit{exploitation}. Each vSC agent applies a procedure shown in Algorithm \ref{alg:algorithmdqn}.

\begin{algorithm}[H]
\caption{DDRL based control }
\begin{algorithmic} 
\State Initialize replay memory $D_i$  of capacity $M$
\State Initialize action-value function, $Q_i$, with random weights $\boldsymbol{\theta}_i$
\For {each episode}:
\State Initialize $X^t_i = (h^t, m^t,  L^t_i, \boldsymbol{B^t})$ - observation from the \\ \hspace{1.3cm} scenario
\For {each step, $t$, of episode}:
\State With probability $\epsilon$ select a random action $A^t$
\State Otherwise $A^t = \max_{A} Q_i (X^t_i, A; \boldsymbol{\theta}_i$)
\State Take action $A^t$, get reward $r_t$ and observe next state \\ \hspace{1.3cm} $X^{t+1}_i$
\State Store transition ( $X_i^t$, $A_i^t$, $r_t$, $X_i^{t+1}$)  in $D_i$
\State Sample a random mini-batch of $K$ transitions \\ \hspace{1.3cm} ($X^j$, $A^j$, $r_j$, $X^{j+1}$) from $D_i$
\State Set target value, $y_j$ = $r_j + \gamma \max_{\hat{A}} Q_i (X^{j+1}, \hat{A} ; \boldsymbol{\theta}_i$)
\State Perform a gradient descent step on \\ \hspace{1.3cm}  ($y_j - Q_i (X^j,A^j ; \boldsymbol{\theta}_i))^2$ with respect to $\boldsymbol{\theta}_i$
\EndFor
\EndFor
\end{algorithmic}
\label{alg:algorithmdqn}
\end{algorithm}
\section {Numerical Results}

\label{sec:results}

\subsection{Simulation Scenario}
According to the traffic model defined in Section~\ref{sec:eh-traffic-profiles}, user activities are categorized as heavy users with an activity of $900$ MB/hr and ordinary users with an activity of $112.5$ MB/hr~\cite{auer2010}. Solar energy traces are generated using the SolarStat tool~\cite{Miozzo2014} for the city of Los Angeles. We have considered the commercial Panasonic N235B as the PV module. These panels have single cell efficiencies as high as 21.1\%, which ranks them among the most efficient solar modules at the time of writing, delivering about 186 $\mathrm{W/m^2}$. The solar panel size and battery capacity are dimensioned based on the criteria that the vSC can be fully recharged on a typical winter day \cite{lces}. The simulation parameters and reference power consumption values are given in Table~\ref{tab:table1}.

\begin{table}[ht]
  \centering
  \caption{Simulation parameters.}
  \begin{tabular}{l c}
   Parameter & Value\\
   \hline \hline
   Transmission power of macro cell (dBm) & 43\\
   Transmission power of vSC (dBm) & 38\\
   Bandwidth (MHz) & 20\\
   MIMO Transmission Mode  & 2x2\\
   UEs per vSC & 90\\
   Heavy users ratio & 0.5\\
   Solar panel size ($\mathrm{m}^2$) & 4.48 \\
   Battery capacity (kWh) & 2\\
   $B_{th}$ & $20\%$\\
   $P_{RF_{vSC}}$ & $2.6$ W\\
   $P_{PA_{vSC}}$ & $71.4$ W\\
   $P_{RF_{MBS}}$ & $9.18$ W\\
   $P_{PA_{MBS}}$ & $1100$ W\\
   GOPS to W conversion factor & 8\\
   $P_{BB_{{static}_{vSC}}}$ & $440$ GOPS\\
   $P_{BB_{{load-dependent}_{vSC}}}$ & $60$ GOPS\\
   $P_{BB_{{static}_{MBS}}}$ & $630$ GOPS\\
   $P_{BB_{{load-dependent}_{MBS}}}$ & $215$ GOPS\\
   $P_{overhead_{vSC}}$ & $0.0\%$\\
   $P_{overhead_{MBS}}$ & $10.0\%$\\
   
  \hline
  \end{tabular}
\label{tab:table1}
\end{table}

We first analyze the behavior of the system when the training is performed off-line. In particular, we considered one year as an episode with time granularity of one hour, since it allows to achieve a correct dimension of the solar power system for cellular BSs, as shown in~\cite{DaSilva2018}. Hence, every hour the agents choose actions corresponding to one of the three possible operative modes, with the goal of minimizing the weighted sum of grid energy consumption and traffic drop rate. 

\subsection{Training Analysis}
As mentioned in Section \ref{sec:ddrl}, the size of the state set for each of the DRL agents is $N+3$, where $N$ denotes the number of vSCs. Important steps prior to feeding these inputs to the neural network of each vSC agent are normalization and transformation, as described in Section \ref{sec:ddrl}. 
Through simulation trials and evaluation, we have selected a $3$-layer dense neural network architecture with $256$,  $128$ and $64$ neurons, respectively. \textit{ReLU} activation function is used for hidden layers while \textit{linear} activation is applied at the output layer. In addition, stochastic gradient descent (SGD) optimizer with momentum and learning rate decay is used for the training. 

For each DRL agent, we have employed an experience replay buffer of size of $M = 2000$ and a mini-batch size of $32$ are chosen for the training. The training procedure for each DRL agent with experience replay is shown in Algorithm \ref{alg:algorithmdqn}.
Given that an episode lasts $1$ year and actions are taken every hour, there are $8640$ time steps within one episode of training. During training, mini-batches of experiences are randomly sampled from the replay buffer to perform forward and back propagation steps in every hour prior to taking actions. After the mini-batch pass, $\epsilon$-greedy policy is applied. Each agent selects a random action (exploration) with probability $\epsilon$, and the action with the maximum Q-value (exploitation) with probability $1-\epsilon$, during training. \dan{In typical RL applications, it is recommended to start with a higher level of exploration and to slowly reduce the exploration rates as the training progresses. The parameters used for training in our DDRL implementation are summarized in Table \ref{tab:table2}. These hyper-parameters are selected through simulation trials, and they correspond to the best values resulting in the convergence of the DRL agents with the presented neural network architecture.}

\begin{table}[ht]
  \centering
  \caption{Training parameters.}
  \begin{tabular}{l c}
   Parameter & Value\\
   \hline \hline
   learning rate ($\alpha$) & 0.01\\
   discount factor ($\gamma$) & 0.9\\
   initial exploration ($\varepsilon$) & 0.9\\
  learning rate decay & 0.01\\
   exploration decay & 0.9\\
   optimizer  & SGD\\
   mini-batch size & 32\\
   number of layers & 3\\
  \hline
  \end{tabular}
\label{tab:table2}
\end{table}

\begin{figure}[h]
\centering
\captionsetup{position=top}
\captionsetup[subfloat]{captionskip=-1pt}
\vspace*{-0.0cm}
 \hspace*{-0.0cm}\subfloat[\label{resi}]{\includegraphics[width=3.2in]{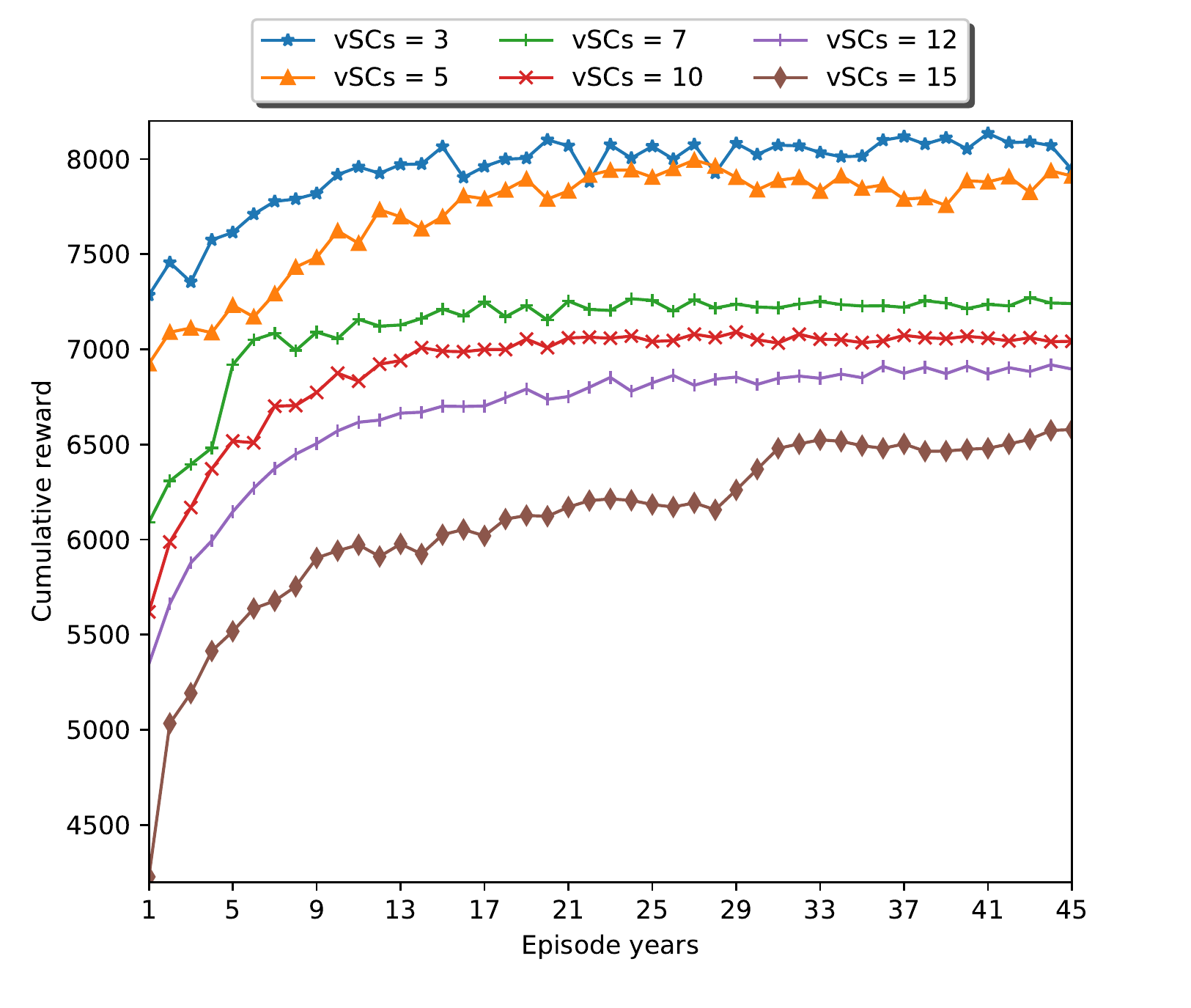}}\hspace{-0.0em}\quad
 \subfloat[\label{offi}]{\includegraphics[width=3.2in]{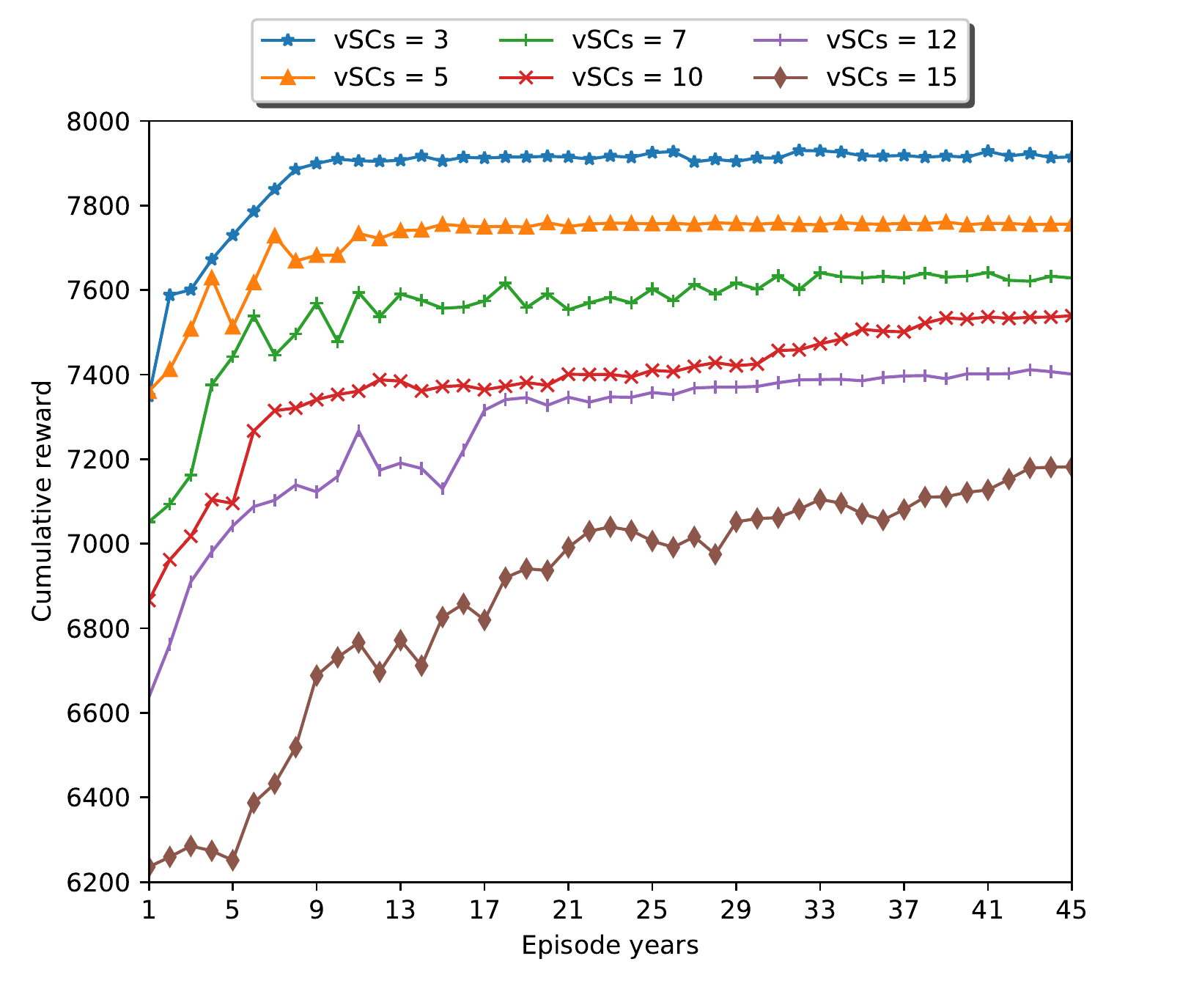}}\hspace{-0.0em}
 \vspace*{-0.0cm}
\caption{Cumulative reward vs number of vSCs: (a)~Residential, and  (b)~Office traffic profiles.}
\label{fig:cr_convergence}
\end{figure}

Cumulative rewards during the training procedure of vSC agents are shown in Figure \ref{fig:cr_convergence} for residential and office area traffic profiles. The training of each DRL agent is performed during $45$ episodes. As it is shown in Figure \ref{fig:cr_convergence}, the scenarios with higher number of vSCs require relatively longer training phase to reach stability and higher rewards. Moreover, the maximum cumulative reward reached at convergence decreases as the number of vSCs increases. This arise mainly due to the non-stationarity of the environment reflecting the conflicts among vSC policies. Finally, we notice that office profile results in a faster training phase and more stable behavior at convergence. In office scenario, traffic and EH phenomena are synchronous and facilitate an easier learning process compared to the residential scenario, in which the higher values of traffic demands appear during evening hours.

\subsection{Policy Characteristics}
\label{sec::Policies}
In this section, we investigate the characteristics of the DDRL policies adopted by the vSCs. In DDRL, each agent chooses its policy in coordination with other agents with a common goal of minimizing the grid energy consumption and the dropped traffic rate simultaneously. As a result, even though the agents' policies can be different, the reward signal each agent acquires is the same, as provided in (\ref{eq:reward}). This helps agents to jointly learn their own policies towards the direction of achieving a system wide goal. In addition, FQL policies are also shown here for comparison. The FQL controller design and training procedures are described in \cite{fss-comcom}. It is important to note that the FQL controller state space consists of the vSC's battery level, energy arrival and vSC's and MBS's traffic load levels. Therefore, agents in FQL are coordinated only via the MBS's traffic load and they are not aware of others vSCs' battery conditions. For the case of $3$ vSCs, both DDRL and FQL solutions are compared against an offline solution based on DP and described in Section \ref{sec:network-model}. Due to the offline model computational complexity, we could not show this comparison for higher number of vSCs.

\begin{figure}[h]
\centering
\captionsetup{position=top}
\captionsetup[subfloat]{captionskip=-1pt}
\vspace*{-0.0cm}
 \hspace*{-0.0cm}\subfloat[\label{offline}]{\includegraphics[width=2.8in]{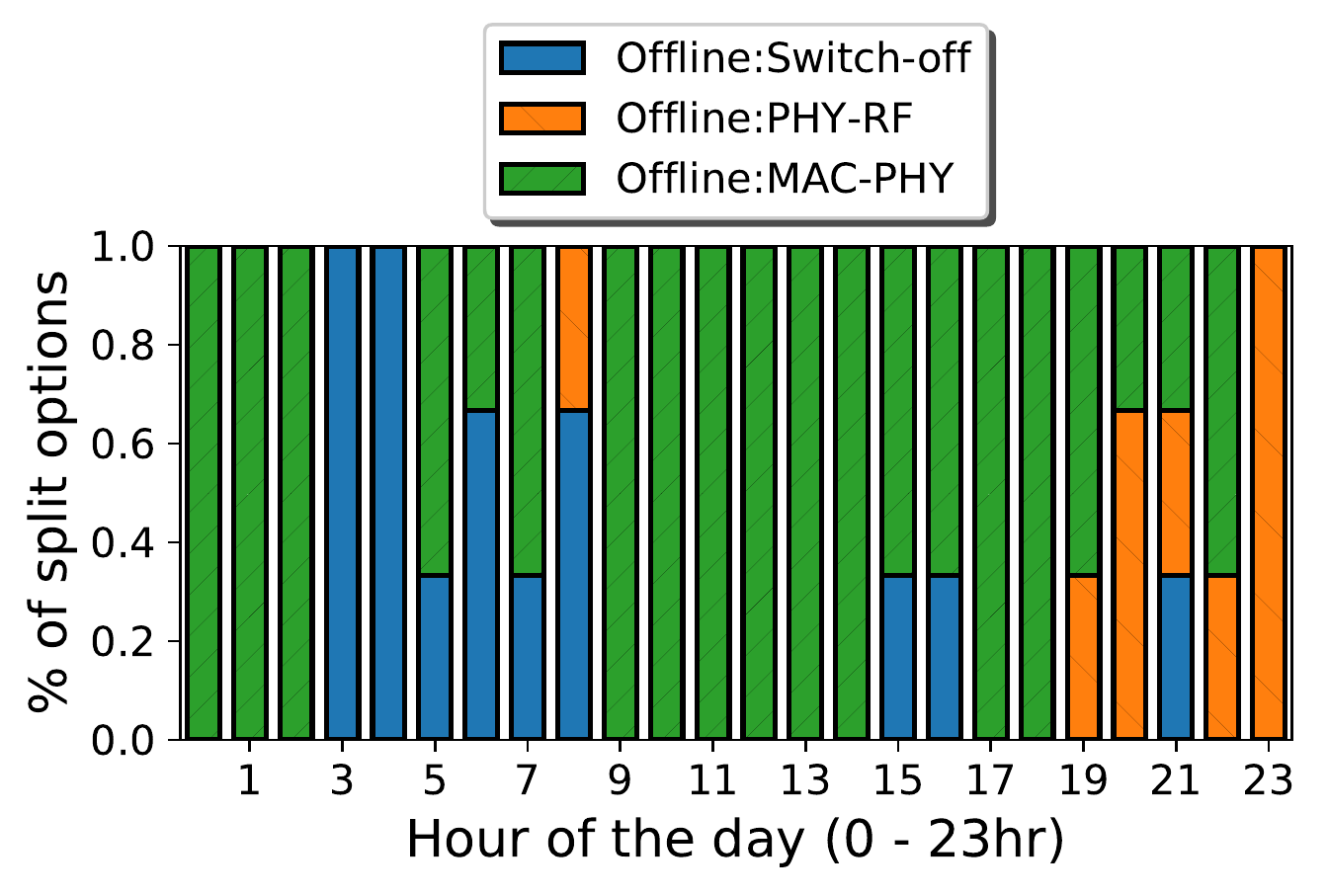}}\hspace{-0.0em}\quad
 \subfloat[\label{fql}]{\includegraphics[width=2.8in]{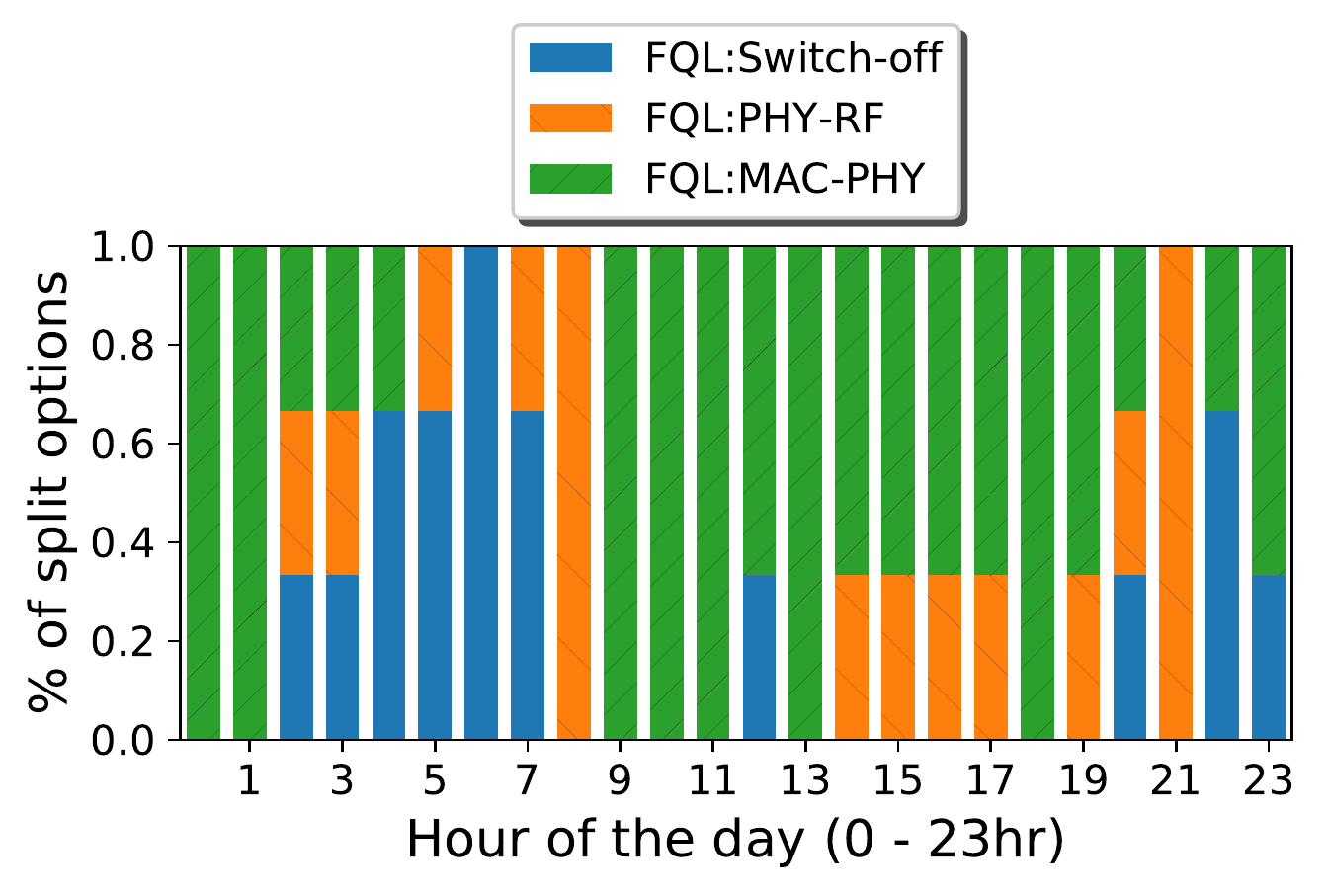}}\hspace{-0.0em}\quad
 \subfloat[\label{ddrl}]{\includegraphics[width=2.8in]{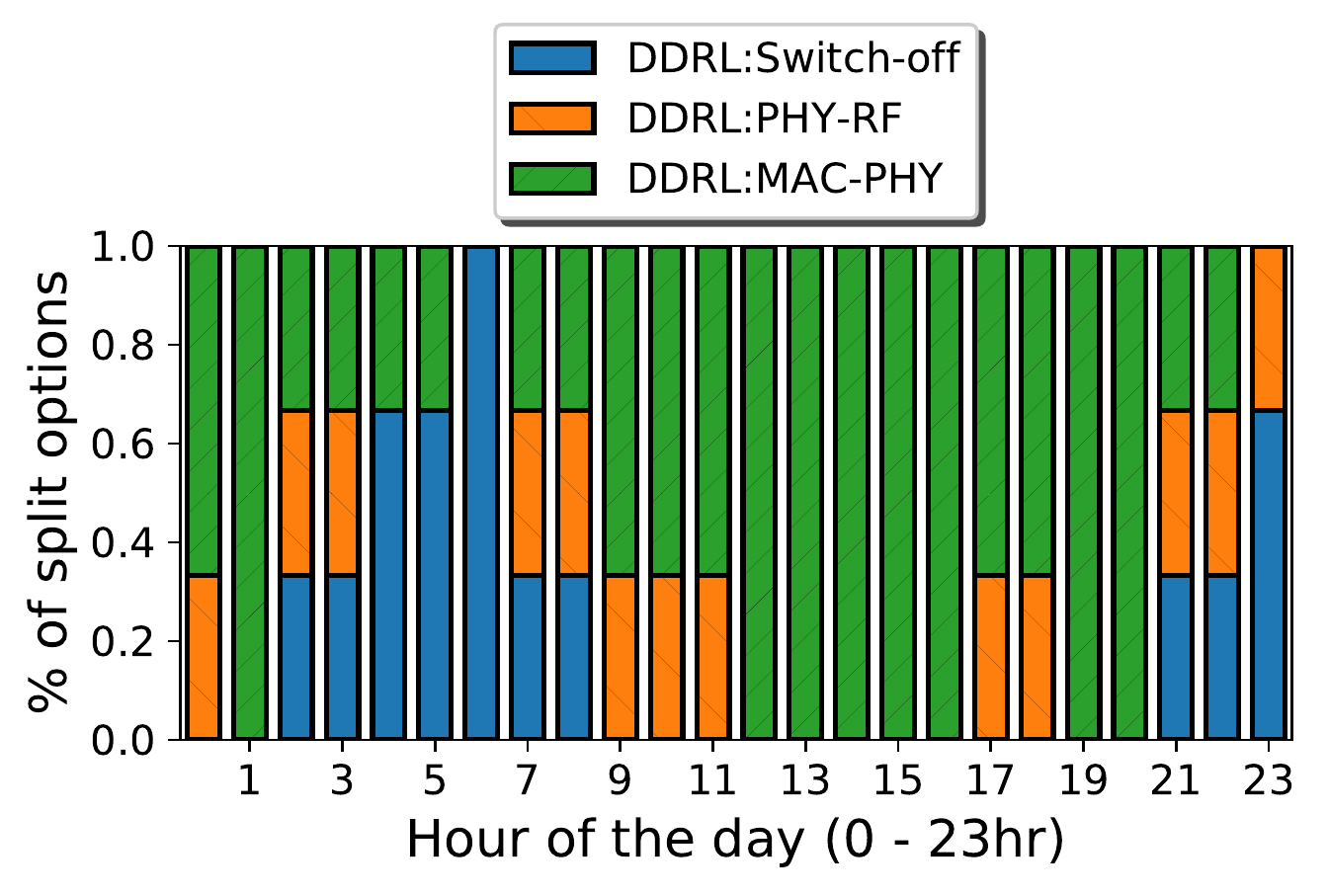}}\hspace{-0.0em}
 \vspace*{-0.25cm}
\caption{Average winter day policies of $3$ vSCs: (a)~Offline (b)~FQL (c)~DDRL.}
\label{fig:winter_nsc3}
\end{figure}

\begin{figure}[h]
\centering
\captionsetup{position=top}
\captionsetup[subfloat]{captionskip=-1pt}
\vspace*{-0.0cm}
 \hspace*{-0.0cm}\subfloat[\label{offline}]{\includegraphics[width=2.8in]{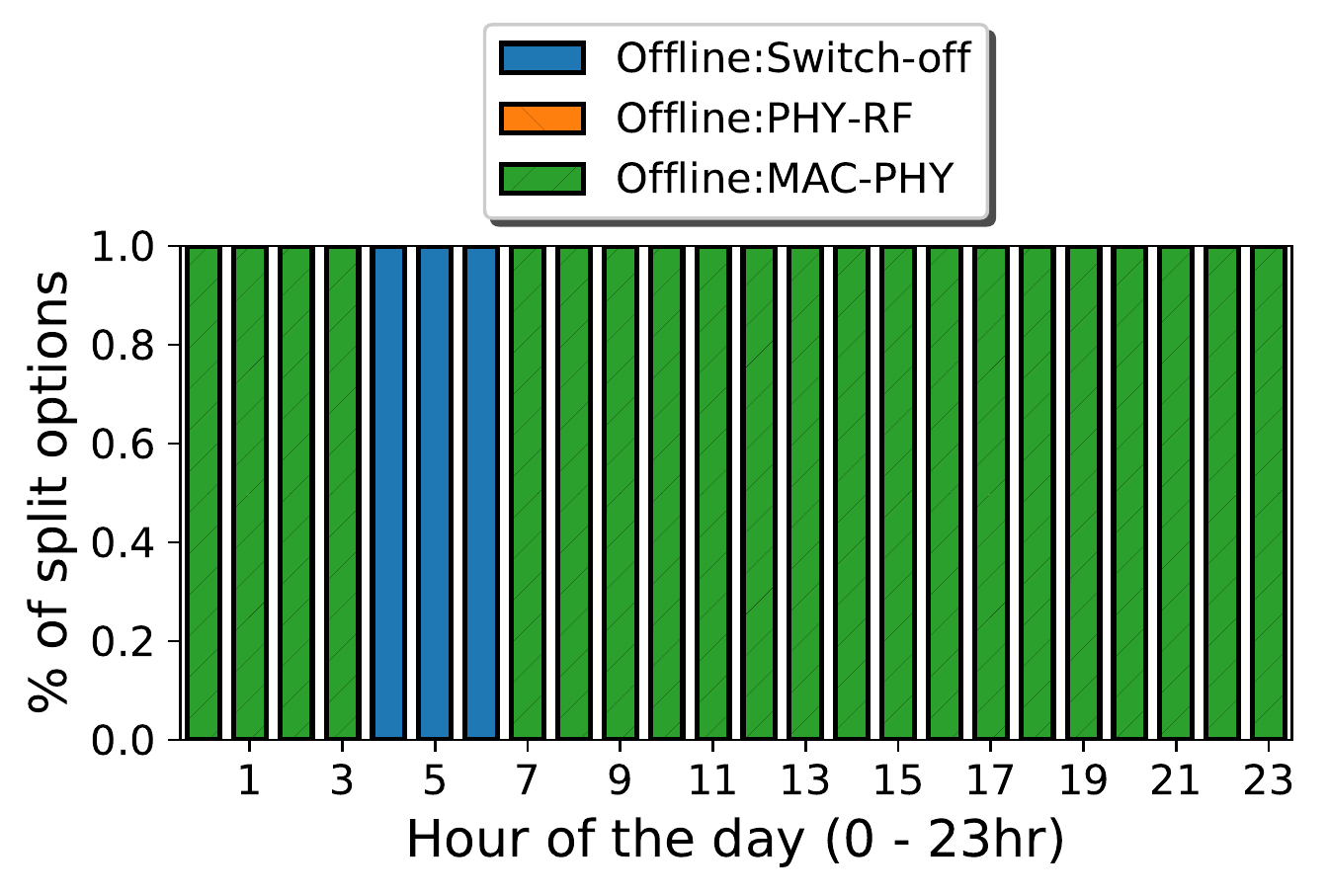}}\hspace{-0.0em}\quad
 \subfloat[\label{fql}]{\includegraphics[width=2.8in]{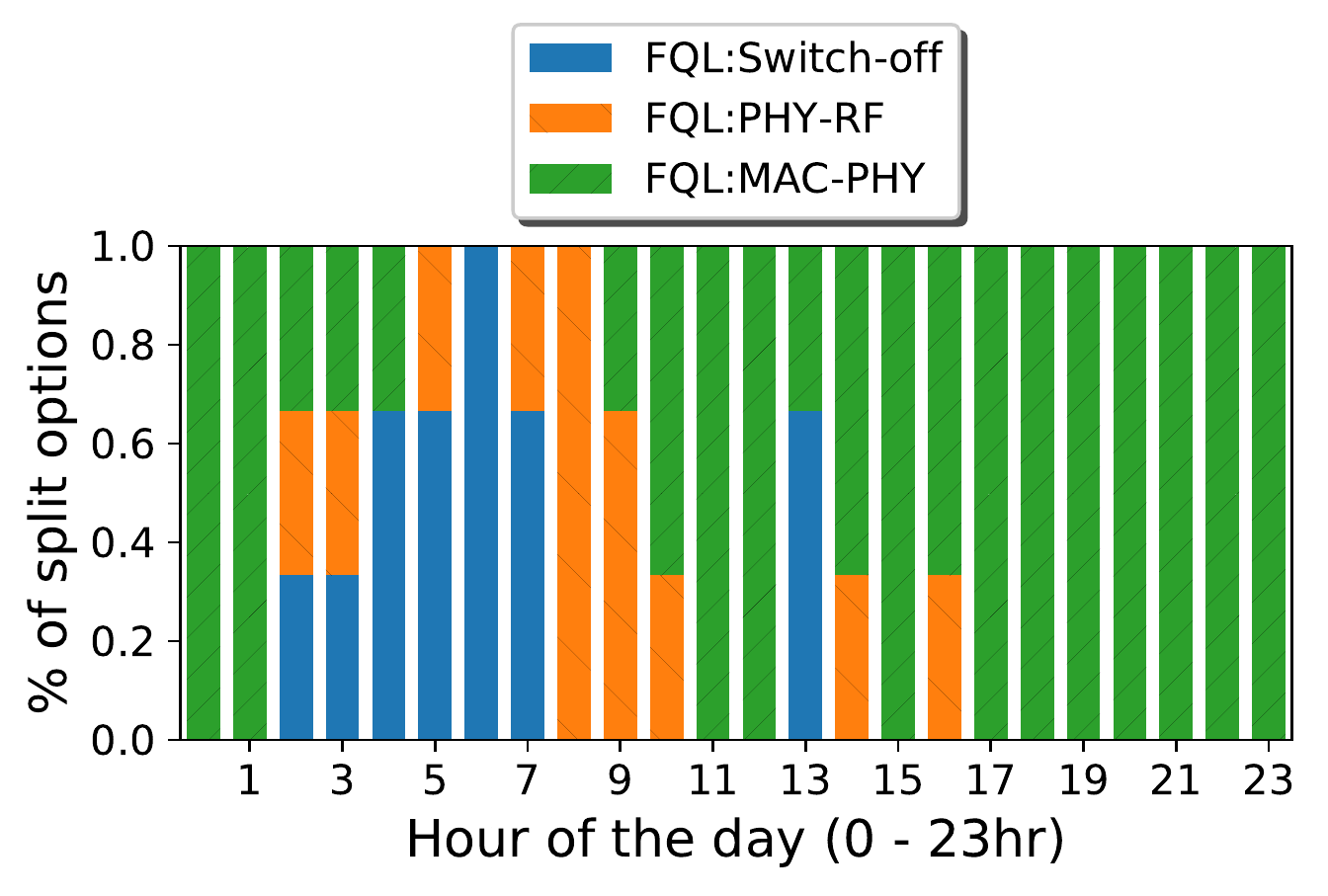}}\hspace{-0.0em}\quad
 \subfloat[\label{ddrl}]{\includegraphics[width=2.8in]{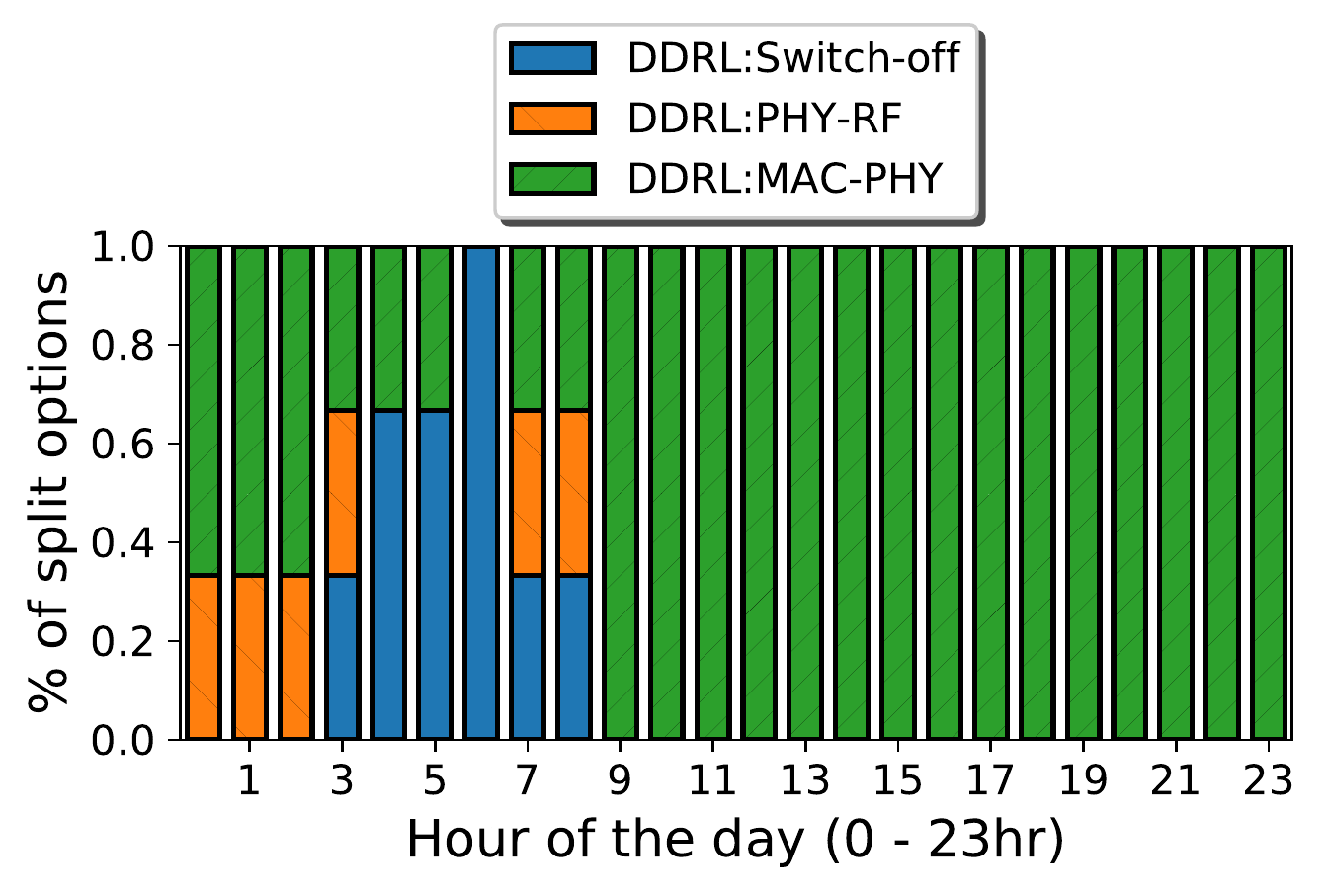}}\hspace{-0.0em}
 \vspace*{-0.25cm}
\caption{Average summer day policies of $3$ vSCs: (a)~Offline (b)~FQL (c)~DDRL.}
\label{fig:summer_nsc3}
\end{figure}

\begin{table*}[h]
  \centering
\captionsetup{justification=centering}
\caption{Average policy characteristics in residential area.}
\scalebox{1}{
\begin{tabular}{|c|c|c|c|c|c|c|}
\hline
 \multirow{2}{*}{\centering number of vSCs} & \multicolumn{3}{|p{6cm}|}{\centering  Winter} & 
    \multicolumn{3}{|p{6cm}|}{\centering Summer} \\
\cline{2-7}
  &Switch-off(\%) & PHY-RF(\%) & MAC-PHY(\%)& Switch-off(\%) & PHY-RF(\%) & MAC-PHY(\%)\\
\hline
3 &  20.8  &18.1 & 61.1 & 13.9 & 8.3 & 77.8\\
\hline
5  & 32.7 & 23.0 & 44.3 & 15.7 & 17.5& 66.8 \\

\hline
7  & 33.9 & 26.9 &  39.2 & 15.2 & 28.0& 56.8\\

\hline
10 & 33.6 & 24.7 &  41.7 & 16.4 &21.5 & 62.1\\

\hline
12  & 38.4 & 20.3 &  41.3 & 18.1 & 23.9& 58.0\\

\hline
15  & 36.7 & 26.2 &  37.1 & 17.6 & 25.1 & 57.3 \\

\hline
\end{tabular}}
\label{tab:policy-char}
\end{table*}

\begin{figure}[h]
\centering
\captionsetup{position=auto}
\captionsetup[subfloat]{captionskip=-1pt}
\vspace*{-0.0cm}
 \hspace{0em}\subfloat[\label{ddrl}]{\includegraphics[width=3.05in]{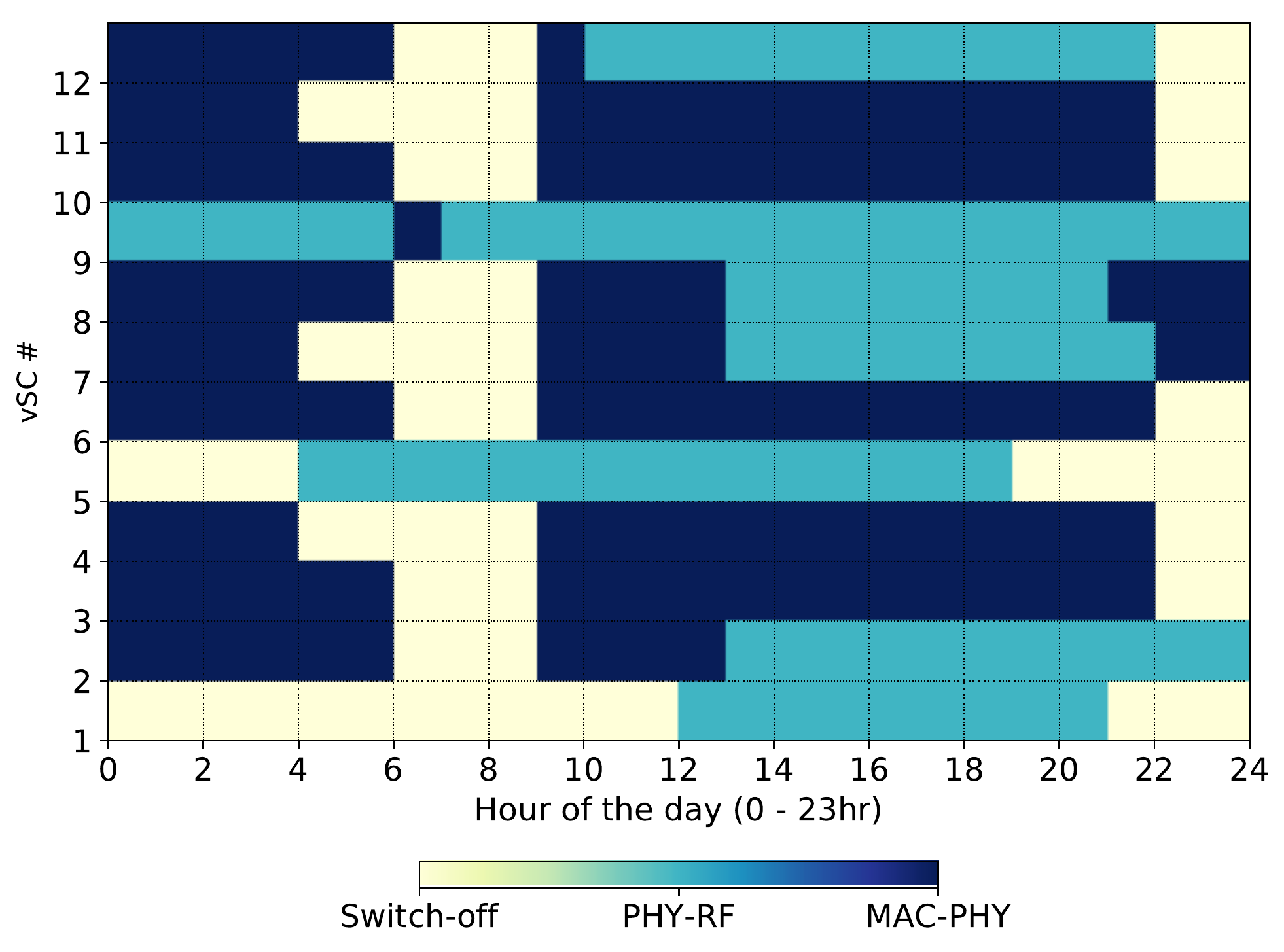}}\quad
 \hspace{0em}\subfloat[\label{fql}]{\includegraphics[width=3.1in]{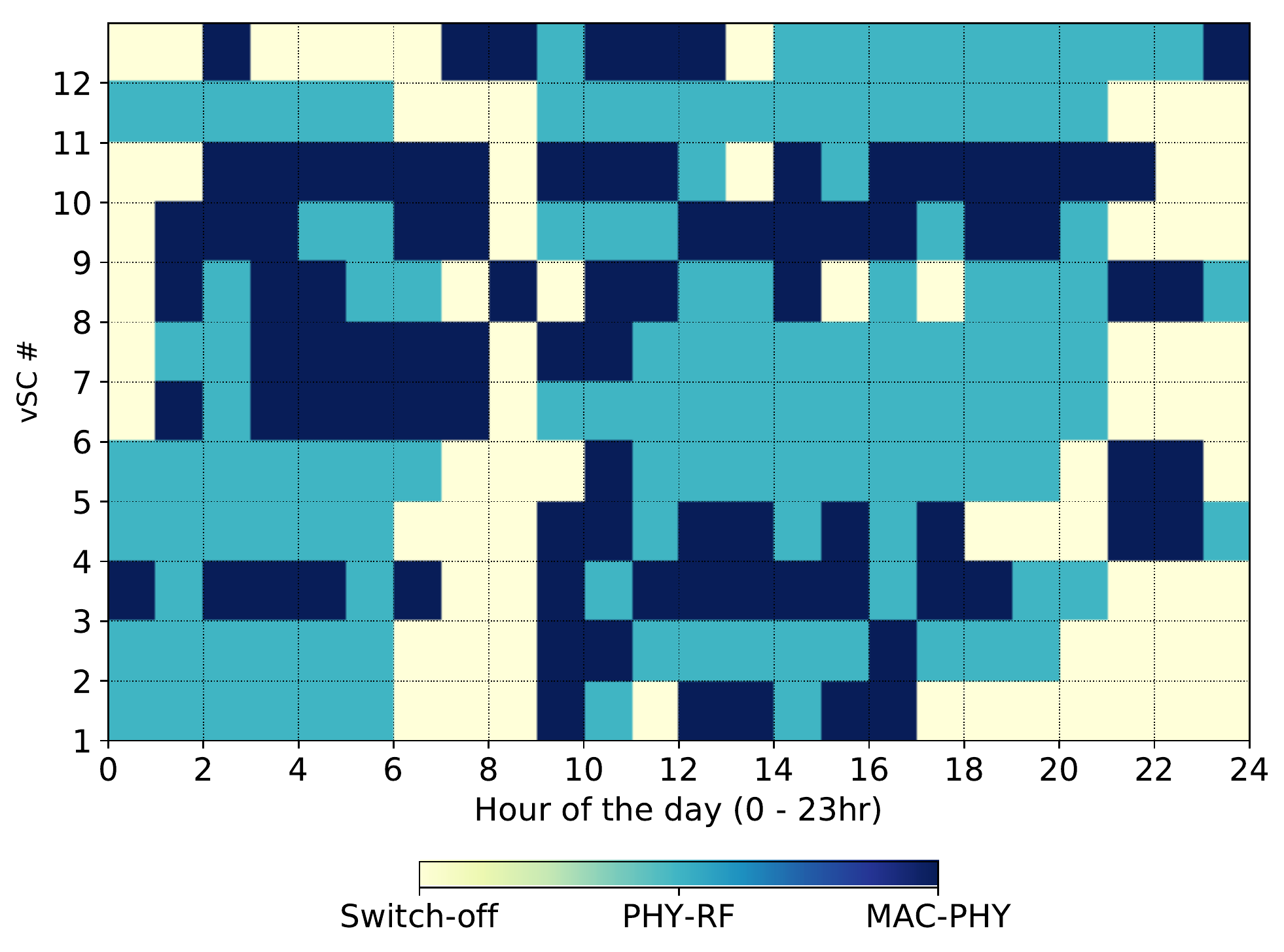}}\quad
 \vspace*{-0.25cm}
\caption{Typical winter day policies with $12$ vSCs: (a)~DDRL (b)~FQL. }
\label{fig:policychar_nsc12_win}
\end{figure}

\begin{figure}[h]
\centering
\captionsetup{position=auto}
\captionsetup[subfloat]{captionskip=-1pt}
\vspace*{-0.0cm}
 \hspace{0em}\subfloat[\label{ddrl}]{\includegraphics[width=3.05in]{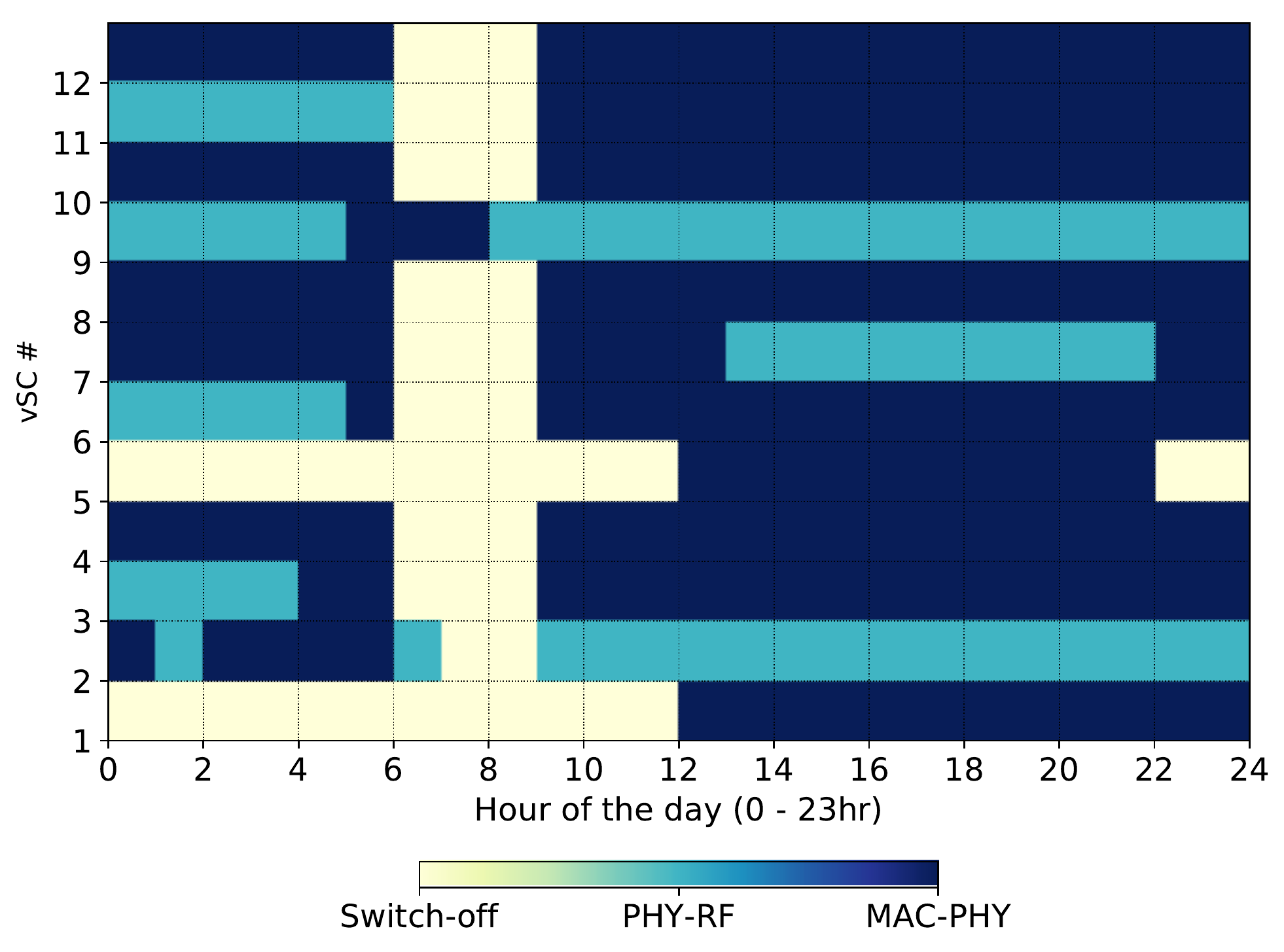}}\quad
 \hspace{0em}\subfloat[\label{fql}]{\includegraphics[width=3.1in]{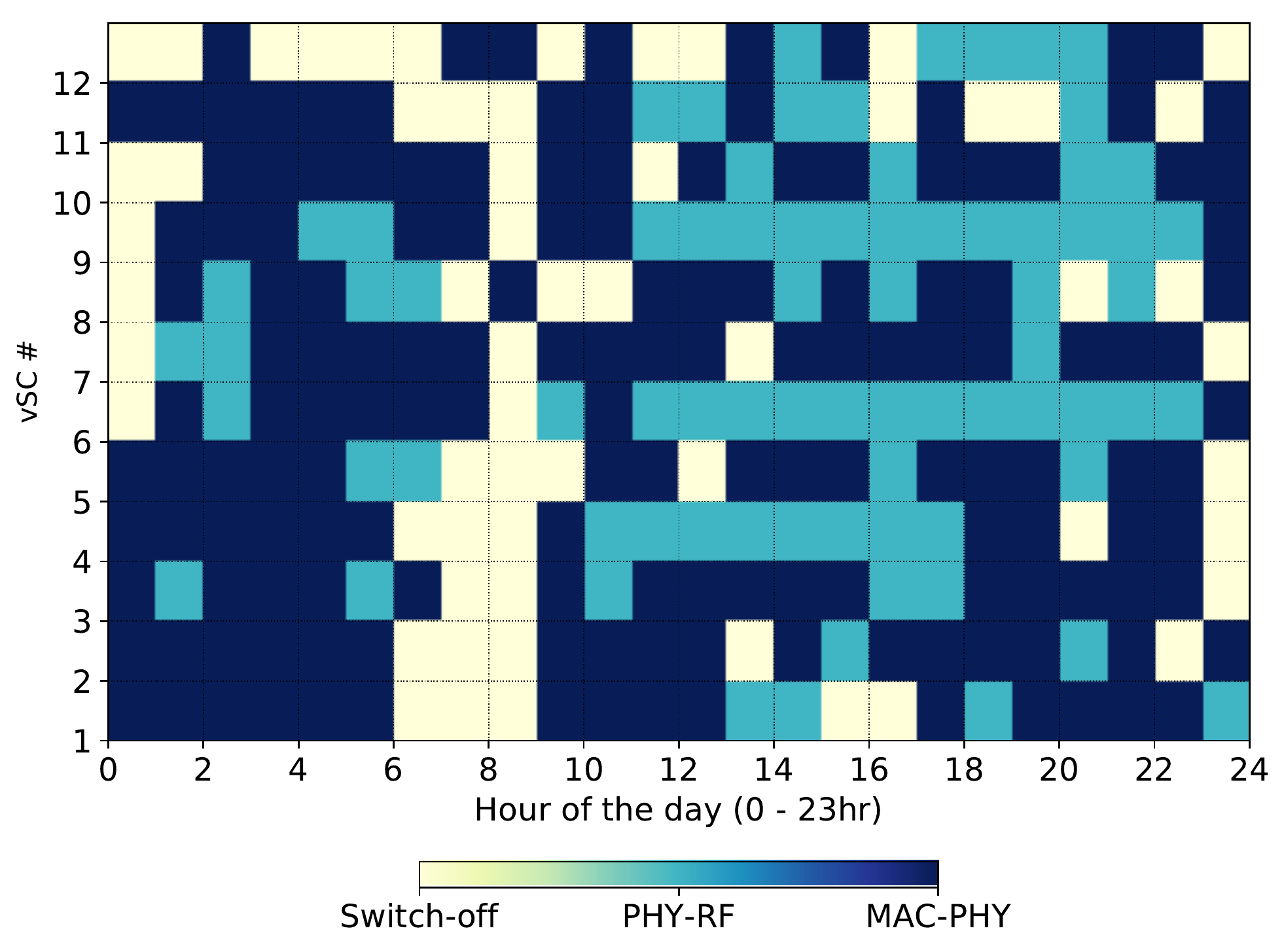}}\quad
 \vspace*{-0.4cm}
\caption{Typical summer day policies with $12$ vSCs: (a)~DDRL (b)~FQL. }
\label{fig:policychar_nsc12_sum}
\end{figure}

The average hourly winter and summer day policies for a scenario of $3$ vSCs in residential area by the offline, FQL and DDRL controllers are shown in Figures \ref{fig:winter_nsc3} and \ref{fig:summer_nsc3}, respectively. \dan{We have selected December for winter, i.e., worst EH, and August for summer, i.e., best EH and the policies observed in each month are averaged hourly to represent average winter and summer daily behavior. It can be observed that DDRL better approximates the offline policy, i.e., the bound, in terms of the selection rates of the three operation modes. For instance, average winter MAC-PHY selection rates were $56\%$, $61\%$ and $67\%$ for the FQL, DDRL and offline policies whereas, in summer, these rates increase to $65\%$, $78\%$ and $87\%$ respectively.} It is interesting to observe from Figure \ref{fig:summer_nsc3} that the offline policy shows no PHY-RF selection during August, thanks to higher energy generation, and indicating the benefit of processing most of the BB functions locally at vSCs. On the other hand, DDRL and FQL result in $8\%$ and $17\%$ PHY-RF selection, respectively, in summer.

\dan{Average winter and summer policies in residential area obtained by DDRL with higher number of vSCs is shown in Table \ref{tab:policy-char}. From Table \ref{tab:policy-char}, an increase in MAC-PHY selection rate and a relative decrease in PHY-RF selection rate is observed in summer with respect to winter policies, whereas, switch-off rate tends to be higher during winter. This shows the adaptation of DDRL policies to higher energy generation, i.e., by selecting MAC-PHY strategy, so that the vSCs execute most of the BB processes locally, in turn saving more grid energy.} On average, summer policies have $20.37\%$ higher MAC-PHY and $18.82\%$ lower switch-off rate than winter day policies.

\dan{An example of daily policies, i.e., without averaging, obtained by DDRL and FQL for a network setup of $12$ vSCs in a residential area are shown in Figures \ref{fig:policychar_nsc12_win} and \ref{fig:policychar_nsc12_sum}, respectively. The figures show that, with both DDRL and FQL, different policies are learned by each vSC characterized by different selection rates of the three operative modes, i.e., PHY-RF, MAC-PHY and switch-off. Moreover, we can see that the policies learned by DDRL and FQL differ. For instance, for a scenario of $12$ vSCs, the winter policies' MAC-PHY selection rates are $41\%$ and $32\%$ by DDRL and FQL, respectively. Whereas, during summer, the MAC-PHY rates increase to $58\%$ and $49\%$, respectively, for DDRL and FQL. The higher MAC-PHY selection rate of DDRL is accompanied by the relatively lower PHY-RF selection. This shows the better flexibility of the DDRL policies according to the seasonal energy inputs.} 
FQL results in less aggressive energy policies for the vSCs via selecting more PHY-RF modes, thereby relying more on the MBS for BB processing, which leads to higher grid energy consumption (as shown in Section \ref{sec:performance}). This is due to the fact that FQL controllers rely only on the normalized MBS traffic load information for coordination rather than the battery states, and hence, it is encouraged to remain switched on in order to reduce the load on MBS and select PHY-RF mode, which is the operative mode with lower energy consumption.

\begin{figure}[h]
\centering
\captionsetup{position=auto}
\captionsetup[subfloat]{captionskip=-1pt}
\vspace*{-0.0cm}
 \hspace{0em}\subfloat[\label{ddrl}]{\includegraphics[width=3.05in]{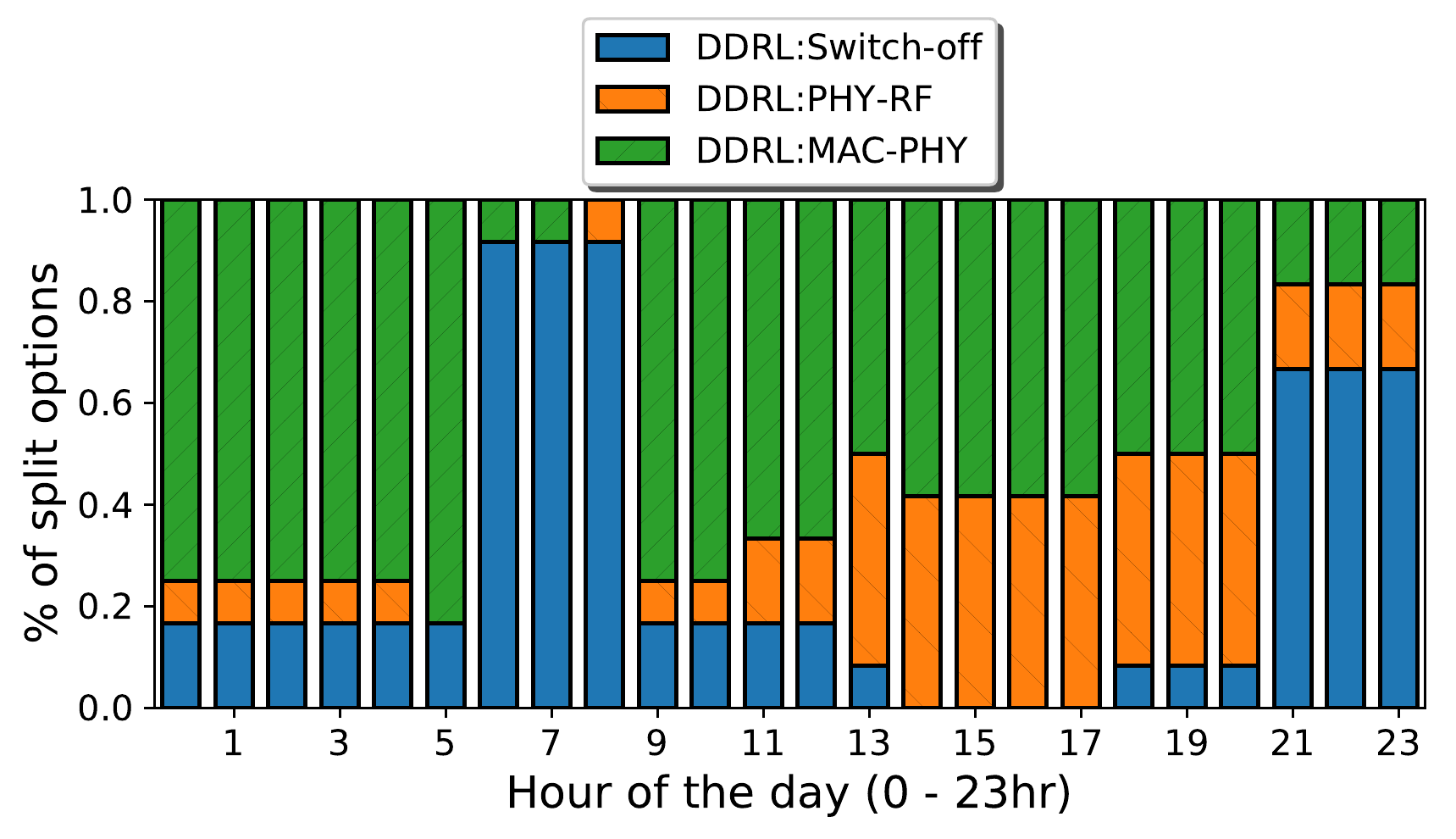}}\quad
 \hspace{0em}\subfloat[\label{fql}]{\includegraphics[width=3.1in]{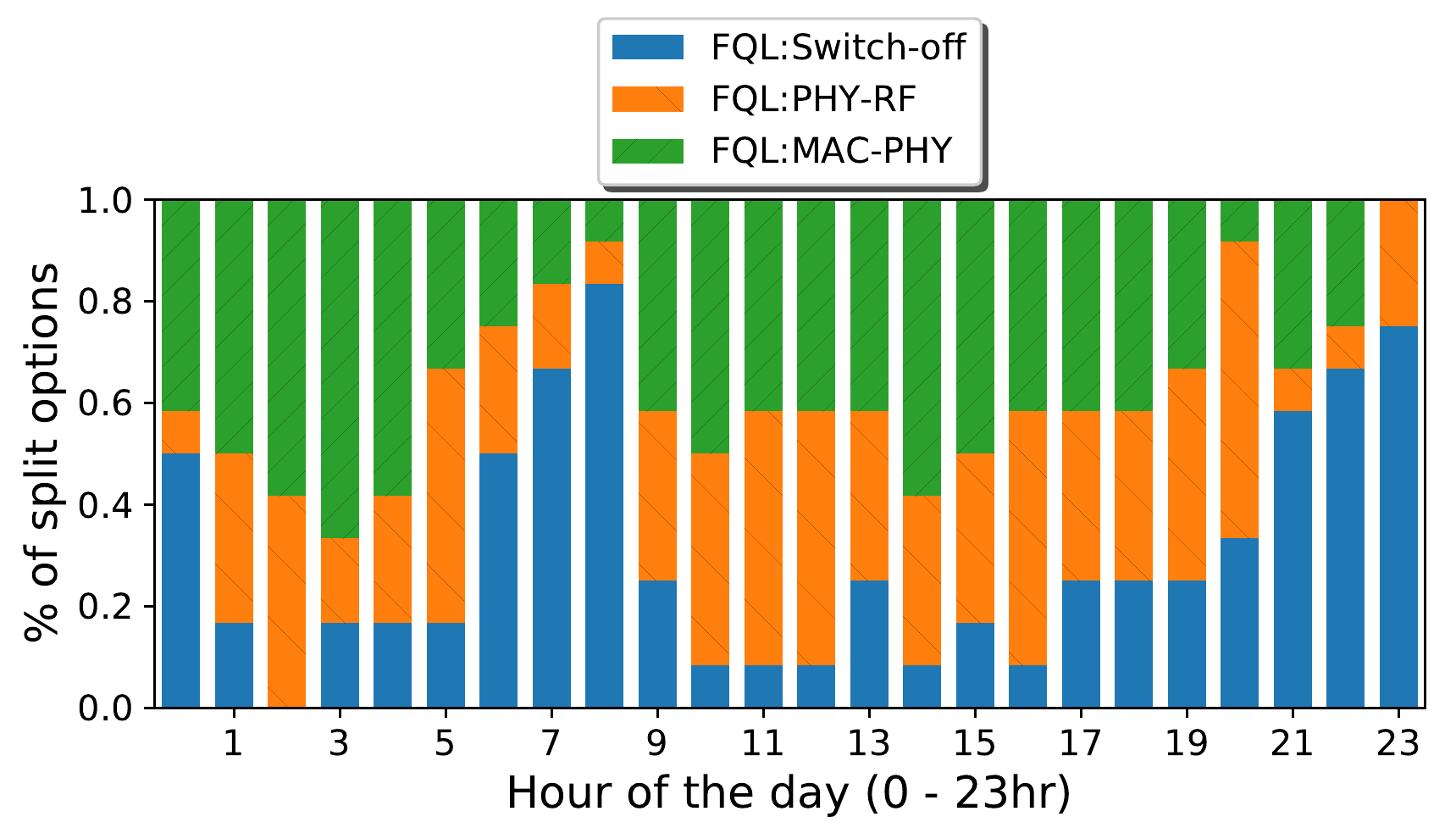}}\quad
 \vspace*{-0.25cm}
\caption{Average winter day policies of $12$ vSCs: (a)~DDRL (b)~FQL. }
\label{fig:winter_bar}
\end{figure}

\begin{figure}[h]
\centering
\captionsetup{position=auto}
\captionsetup[subfloat]{captionskip=-1pt}
\vspace*{-0.0cm}
 \hspace{0em}\subfloat[\label{ddrl}]{\includegraphics[width=3.05in]{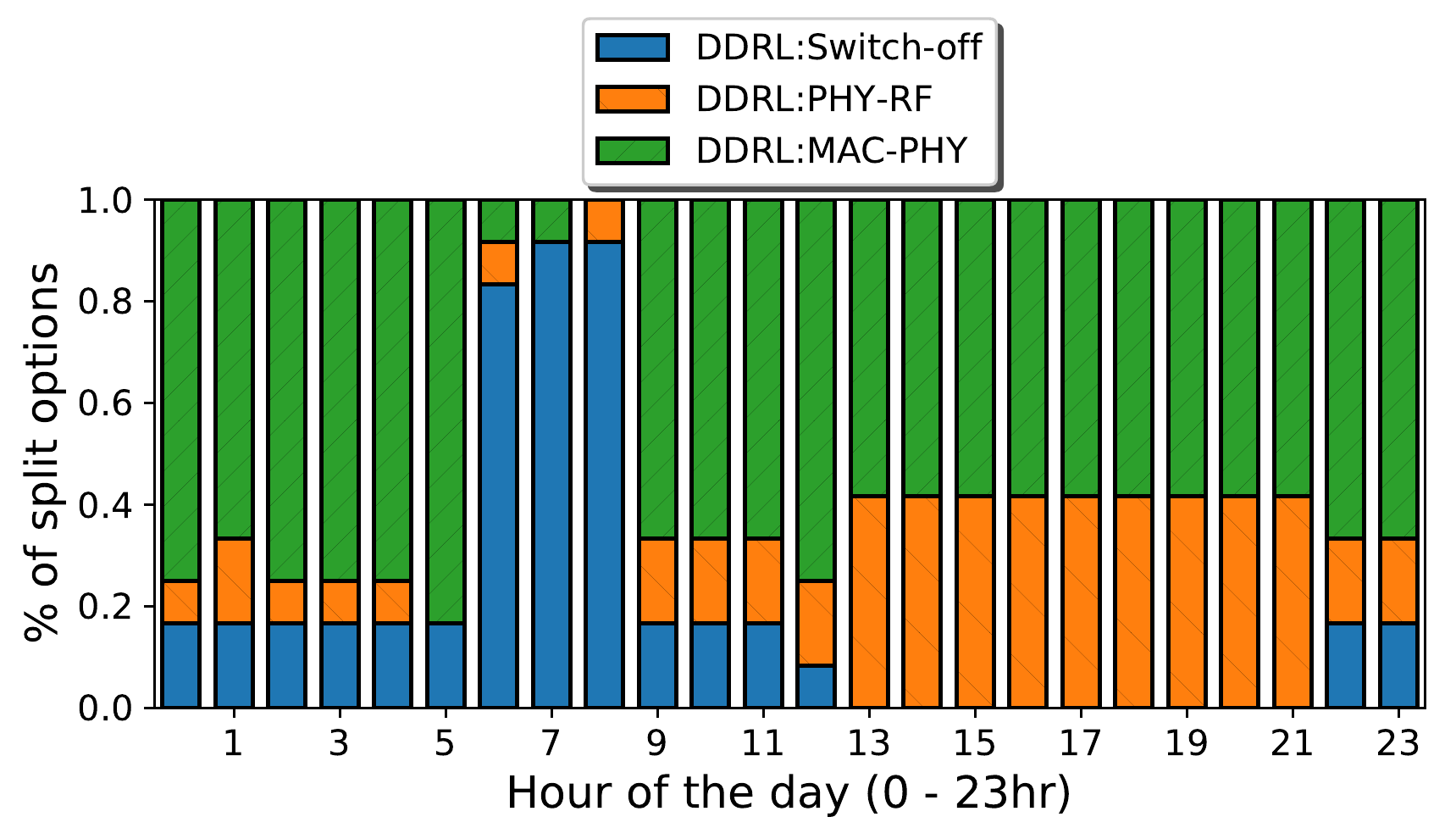}}\quad
 \hspace{0em}\subfloat[\label{fql}]{\includegraphics[width=3.1in]{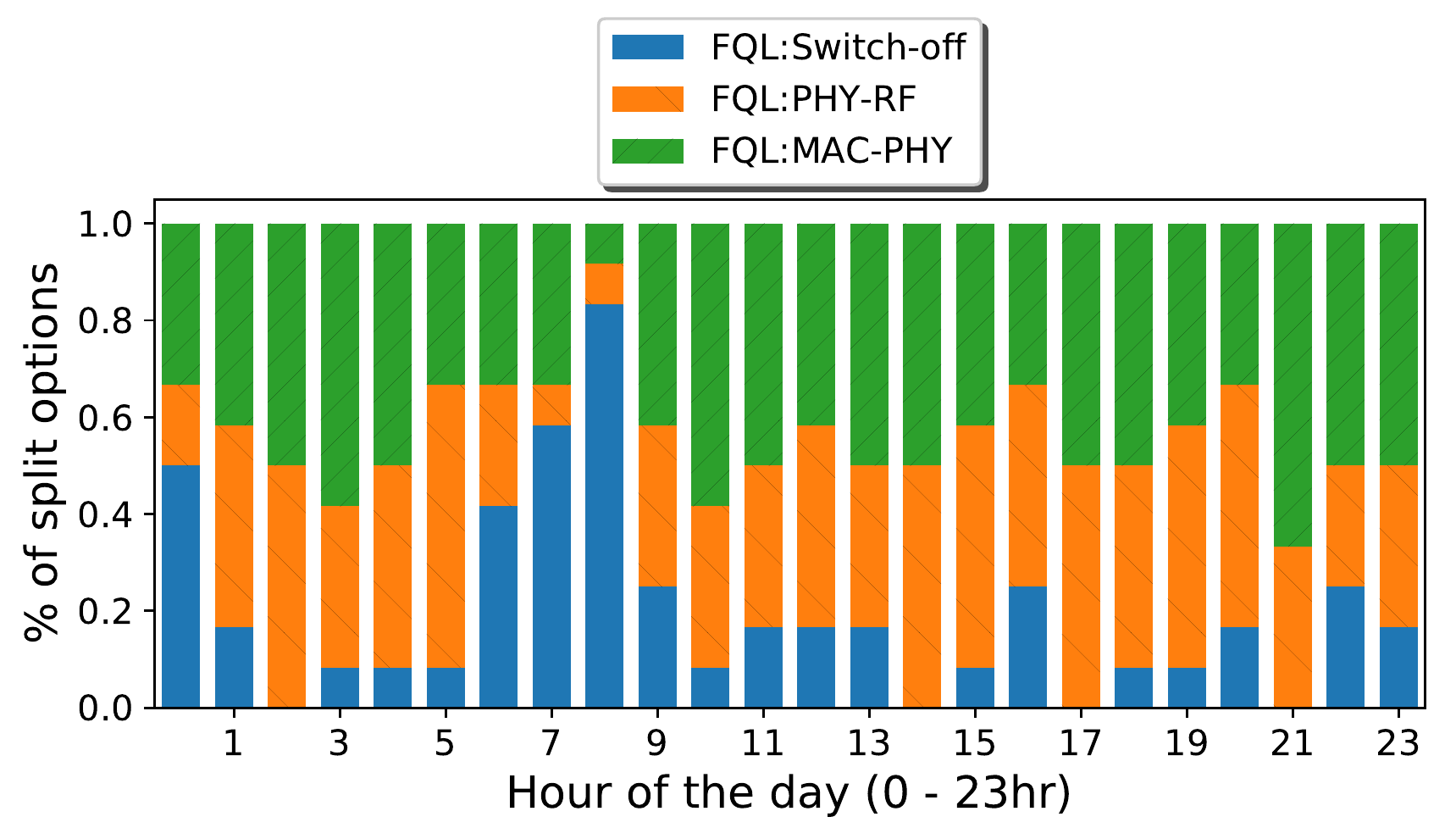}}\quad
 \vspace*{-0.25cm}
\caption{Average summer day policies of $12$ vSCs: (a)~DDRL (b)~FQL. }
\label{fig:summer_bar}
\end{figure}

\dan{Moreover, from the policy characteristics shown in Figures \ref{fig:policychar_nsc12_win} and \ref{fig:policychar_nsc12_sum}, DDRL policies tend to have more stable behavior, i.e., vSCs prefer to stay in a single operative mode for a longer duration than the FQL agents. To better illustrate this behavior, we have shown the average day policies of FQL and DDRL during winter and summer months in Figures \ref{fig:winter_bar} and \ref{fig:summer_bar}, respectively, for a scenario of $12$ vSCs. These behaviors are computed by averaging the vSC policies observed in each day during December and August. As shown in Figures  \ref{fig:winter_bar} and \ref{fig:summer_bar}, the DDRL results in policies that exhibit similar behavior within certain time-slots, whereas FQL policies are characterized by relatively higher variation within a day. The relative stability of DDRL policies is advantageous in limiting the overhead in the SDN/NFV framework for moving the virtual network functions as well as to reduce the frequency of operative mode changes in a deployed infrastructure. }
 
\subsection{Network Performance}
\label{sec:performance}

In this section, we evaluate the performance of the DDRL controller in terms of annual network grid energy consumption and average traffic drop rate. As a comparison benchmark, we also present the performance of FQL policies. The FQL solutions rely on broadcasting the normalized MBS traffic load for coordination. As shown in \cite{fss-comcom}, this broadcasting results in $3.81 - 9.77\%$ grid energy savings as well as $0.3 - 2.21\%$ reduction in annual traffic drop rate with respect to FQL solutions without the normalized MBS load information. This highlights the importance of knowledge exchange for coordinated learning in multi-agent scenarios. Here, we compare the performance of DDRL, where the coordination is achieved through the exchange of battery states, with the FQL solution coordinated via the MBS's traffic load.  In addition, for the case of $3$ vSCs, the network performance of both DDRL and FQL are evaluated against the offline bound studied in \cite{fss-vtc}.

\begin{figure}[h]
	\centering
	\hspace*{0.0cm}\includegraphics[scale=0.4]{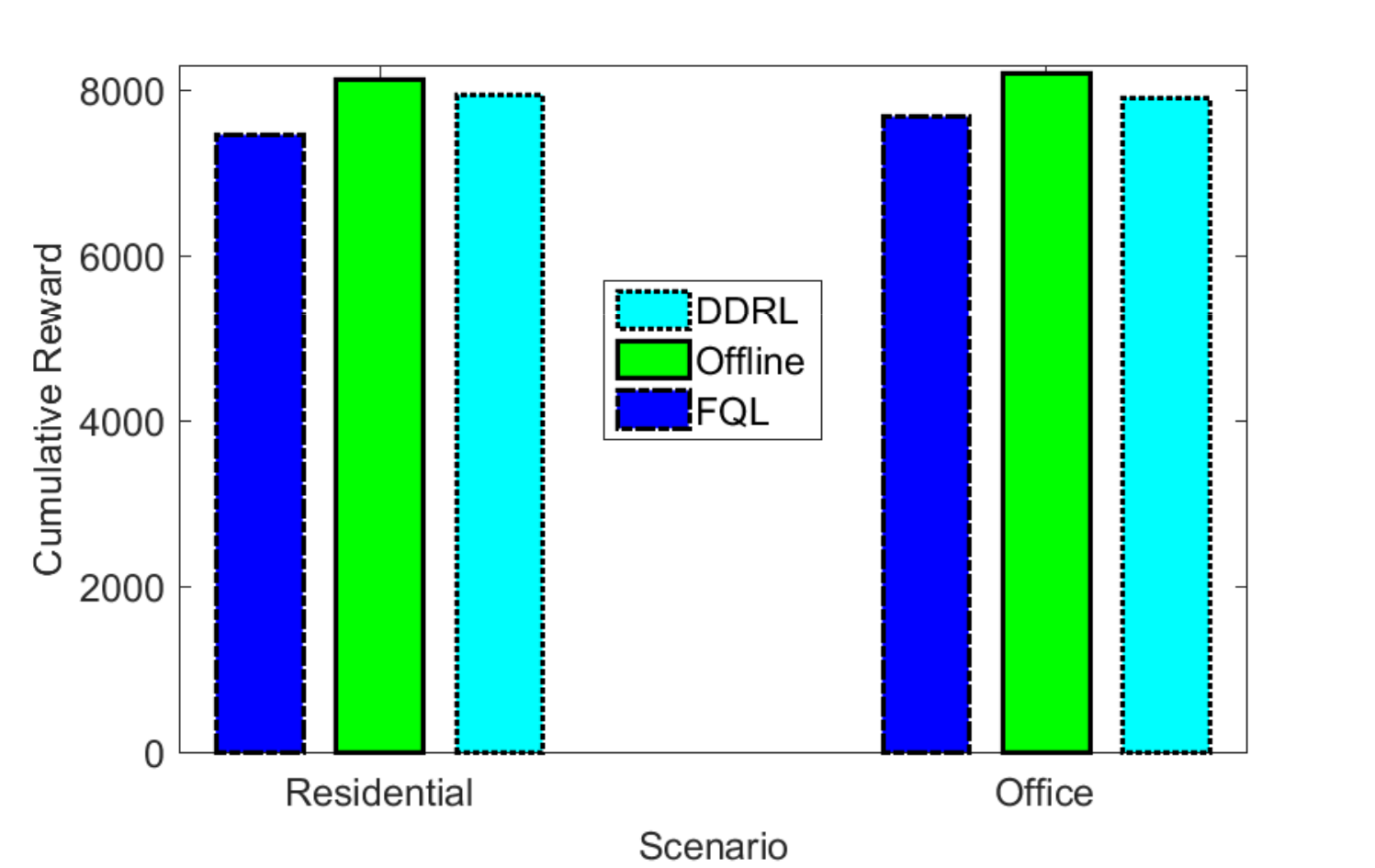}
	\caption{Cumulative reward comparison among the polices obtained by offline optimization, DDRL and FQL.}
	\label{fig:cr_offline_bar}
\end{figure}

Table \ref{tab:bound_RL} shows the performance of DDRL and FQL polices compared with the offline bound of $3$ vSCs. Both DDRL and FQL polices perform close to the offline bound in both residential and office scenarios with only $1.4 - 2.1\%$ and $4.8 - 5.3\%$ increase in annual grid energy consumption, respectively. DDRL performs closer to the offline bound and this can be confirmed by the cumulative reward plot shown in Figure \ref{fig:cr_offline_bar}, where it can be seen that DDRL accumulates up to $97\%$ of the rewards obtained by the offline policy whereas FQL accumulates up to $94\%$. 
\begin{table}[H]
  \centering
\caption{Comparison with the offline bound for $3$ vSCs.}
\scalebox{1}{
\begin{tabular}{|c|c|c|c|c|}
\hline
\multirow{2}{*}{Algorithm} & \multicolumn{2}{|p{3cm}|}{\centering  Grid energy consumption (KWh)} & 
    \multicolumn{2}{|p{2cm}|}{\centering Average drop rate (\%)} \\
\cline{2-5}
 & Residential & Office& Residential & Office \\
\hline
Offline&6775&6712 & 0.0 & 0.0\\

DDRL & 6874 (+1.4\%) & 6857 (+2.1\%)& 0.0 & 0.0\\

FQL &  7136 (+5.3\%)& 7037 (+4.8\%) & 0.0 & 0.0 \\
\hline
\end{tabular}}
\label{tab:bound_RL}
\end{table}

\begin{figure}[h]
\centering
\captionsetup{position=auto}
\captionsetup[subfloat]{captionskip=-0pt}
\vspace*{-0.0cm}
 \hspace{0em}\subfloat[\label{energy_comp}]{\includegraphics[width=2.95in]{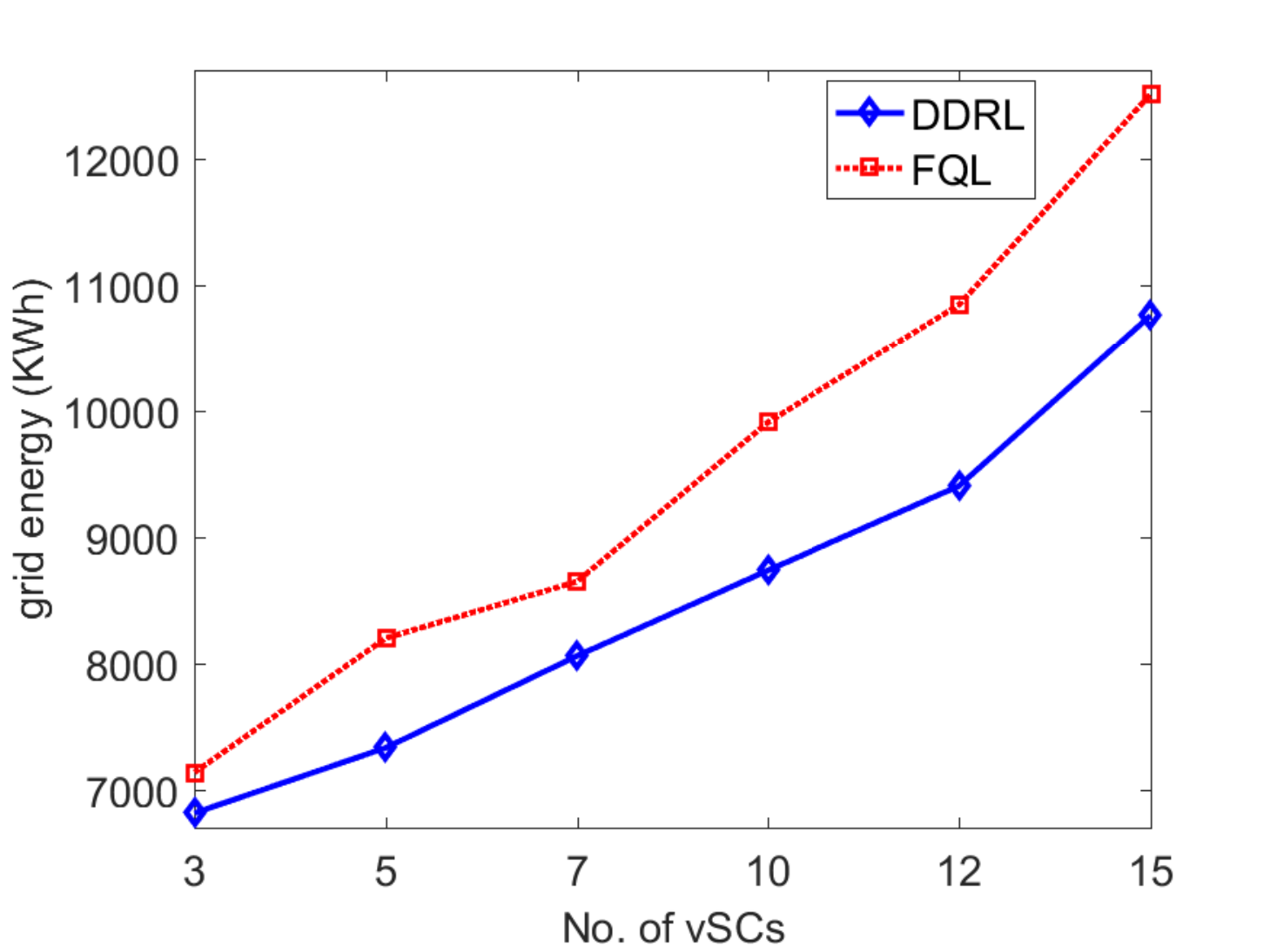}}\quad
 \hspace{0em}\subfloat[\label{drop_comp}]{\includegraphics[width=2.95in]{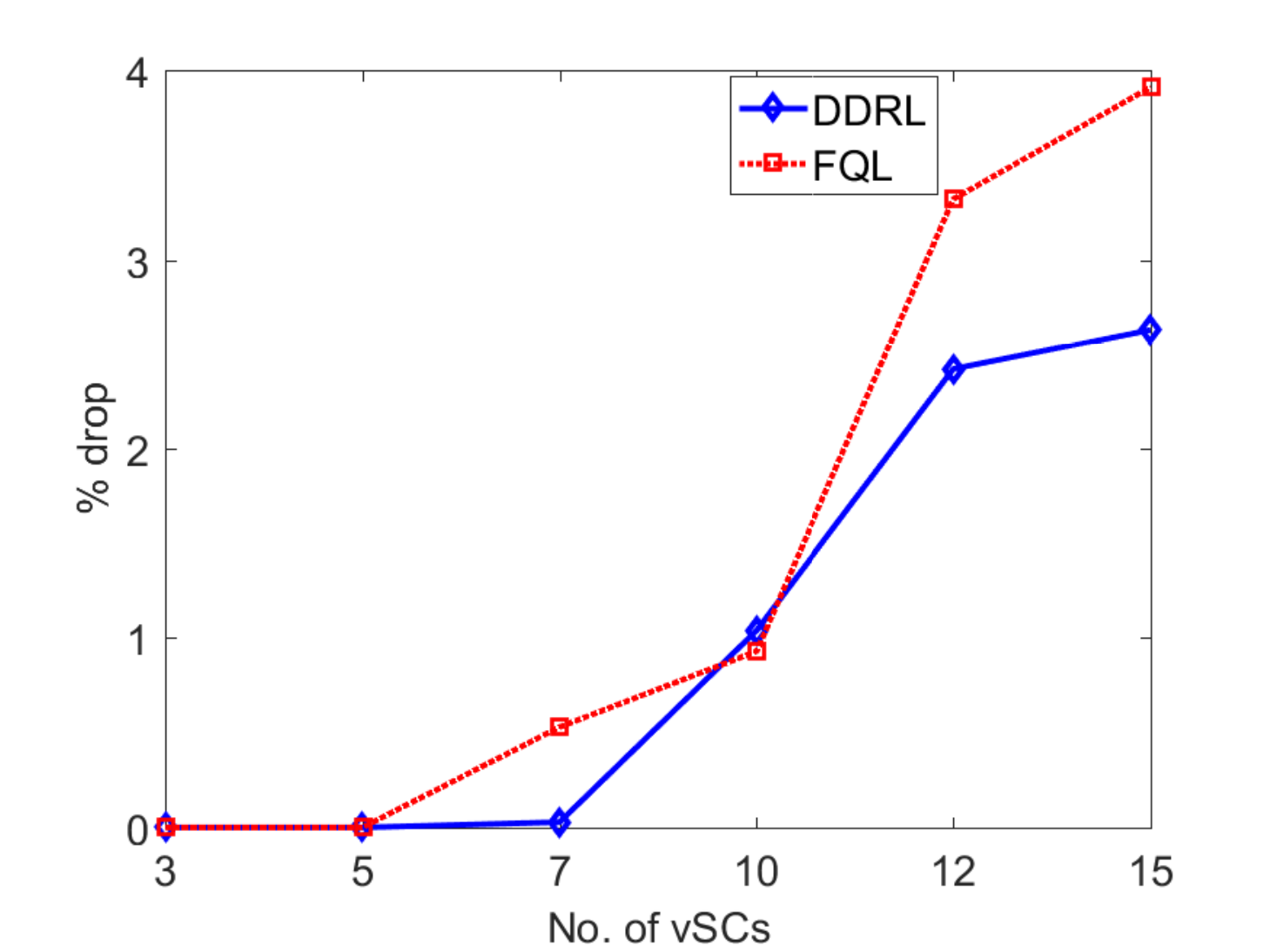}}\quad
 \vspace*{-0.1cm}
\caption{Network performance comparison between DDRL and FQL in residential profile: (a)~Grid energy consumption (KWh) (b)~Average drop rate (\%). }
\label{fig:DDRL-FQL-resi}
\end{figure}

\begin{figure}[h]
\centering
\captionsetup{position=auto}
\captionsetup[subfloat]{captionskip=-0pt}
\vspace*{-0.0cm}
 \hspace{0em}\subfloat[\label{energy_comp}]{\includegraphics[width=2.95in]{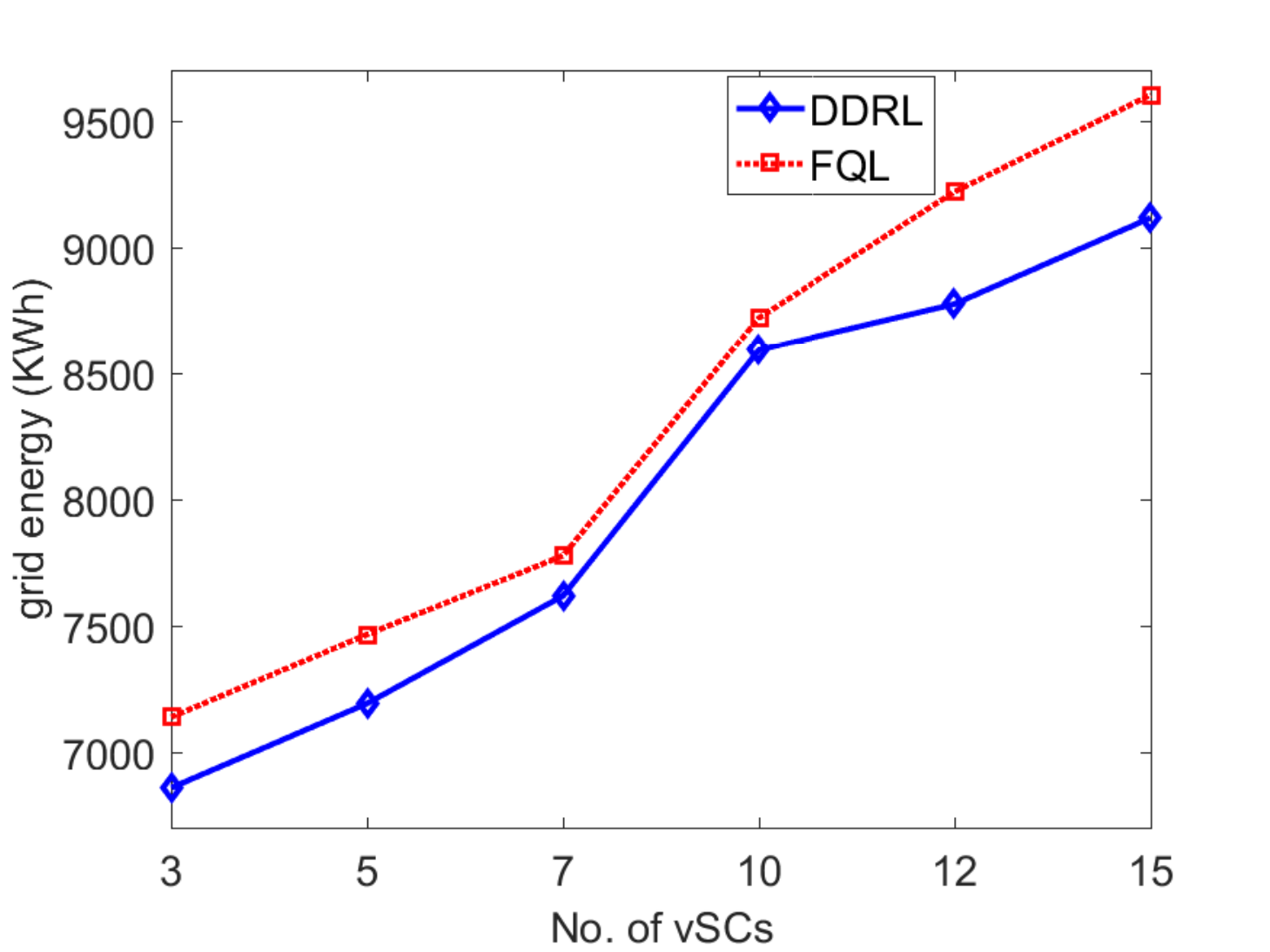}}\quad
 \hspace{0em}\subfloat[\label{drop_comp}]{\includegraphics[width=2.95in]{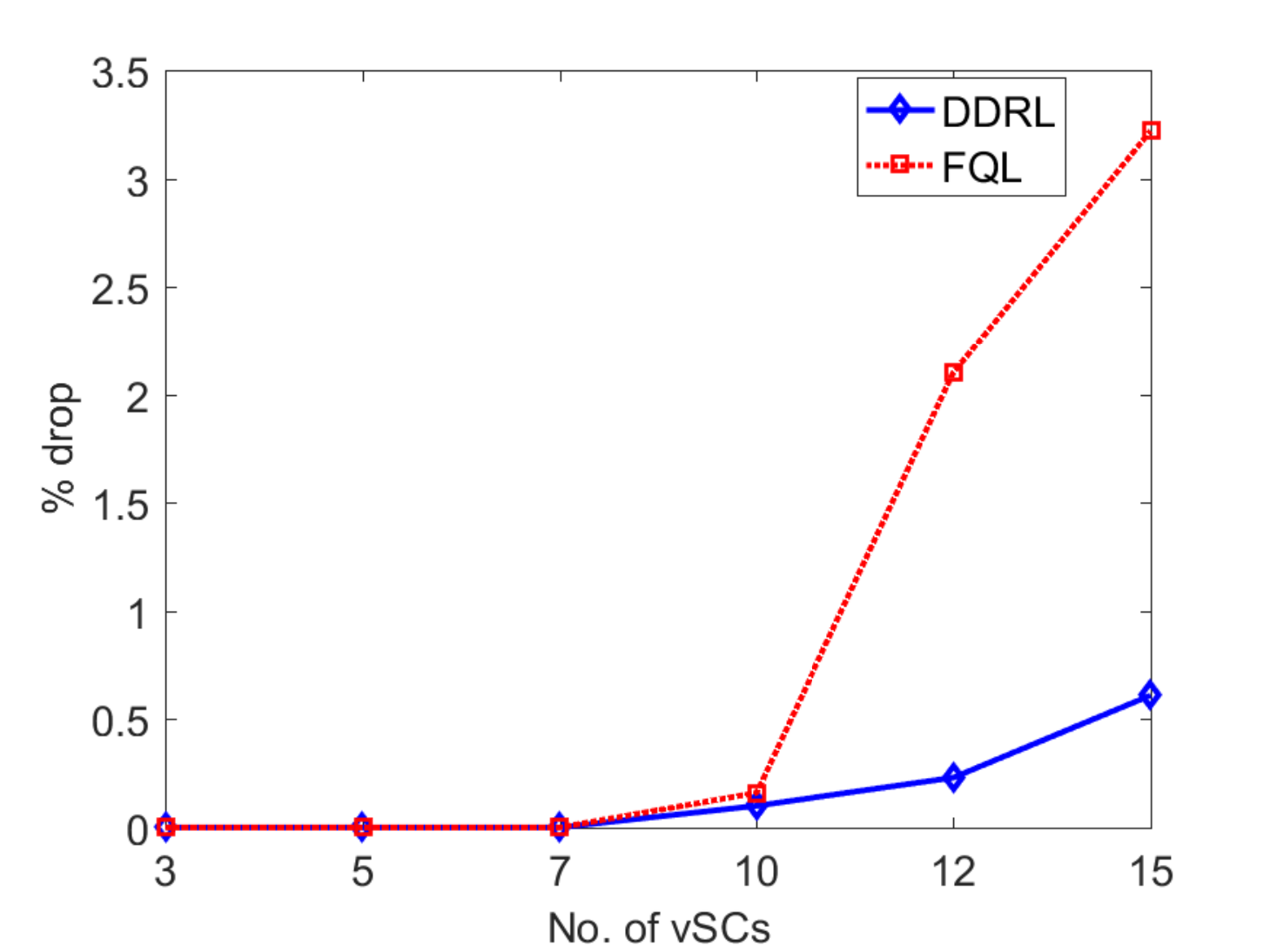}}\quad
 \vspace*{-0.1cm}
\caption{Network performance comparison between DDRL and FQL in office profile: (a)~Grid energy consumption (KWh) (b)~Average drop rate (\%). }
\label{fig:DDRL-FQL-offi}
\end{figure}

\begin{figure}[h]
\centering
\captionsetup{position=top}
\captionsetup[subfloat]{captionskip=-1pt}
\vspace*{-0.0cm}
 \hspace*{-0.0cm}\subfloat[\label{offline}]{\includegraphics[width=3in]{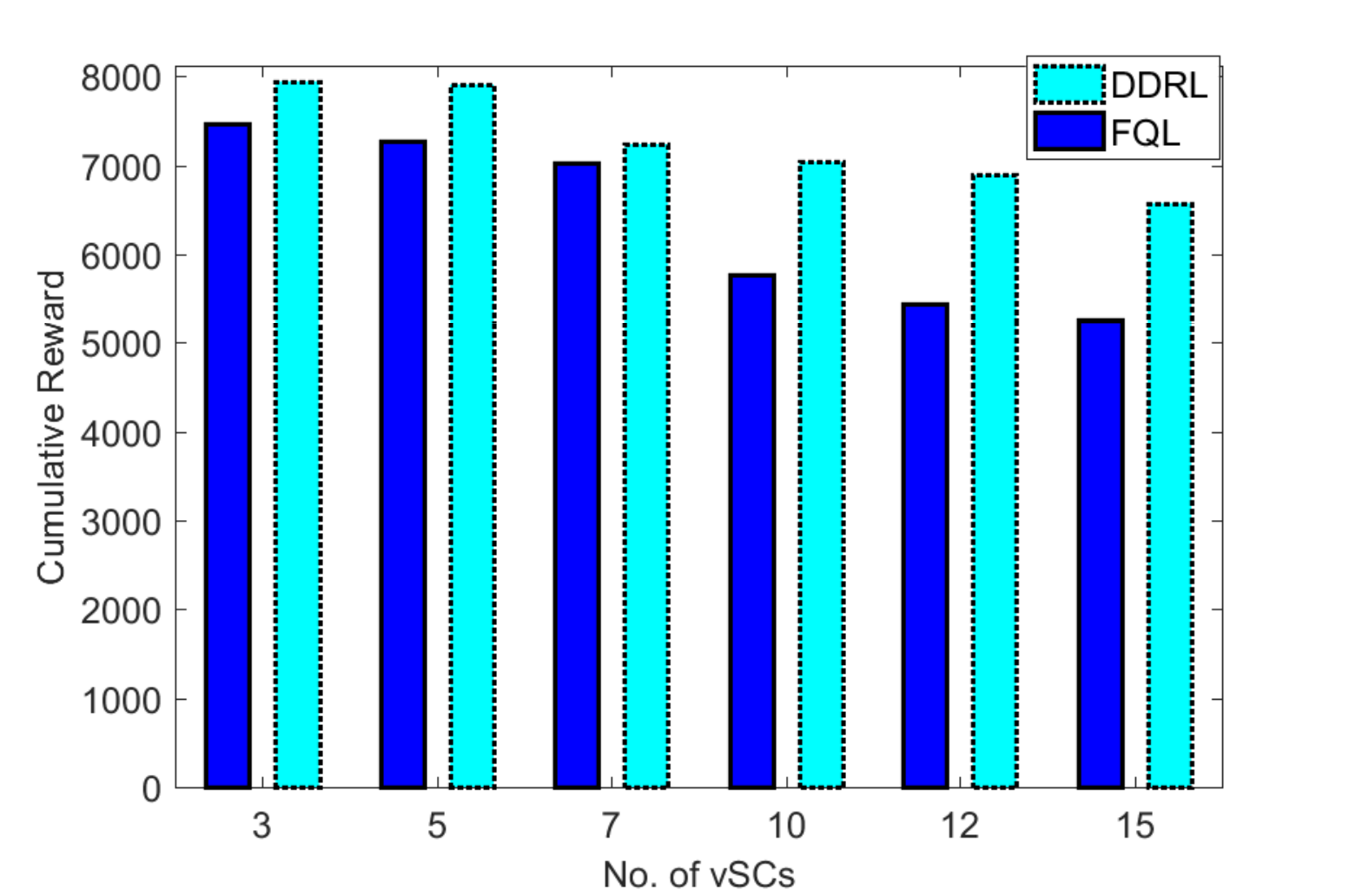}}\hspace{-0.0em}\quad
 \subfloat[\label{ddrl}]{\includegraphics[width=3in]{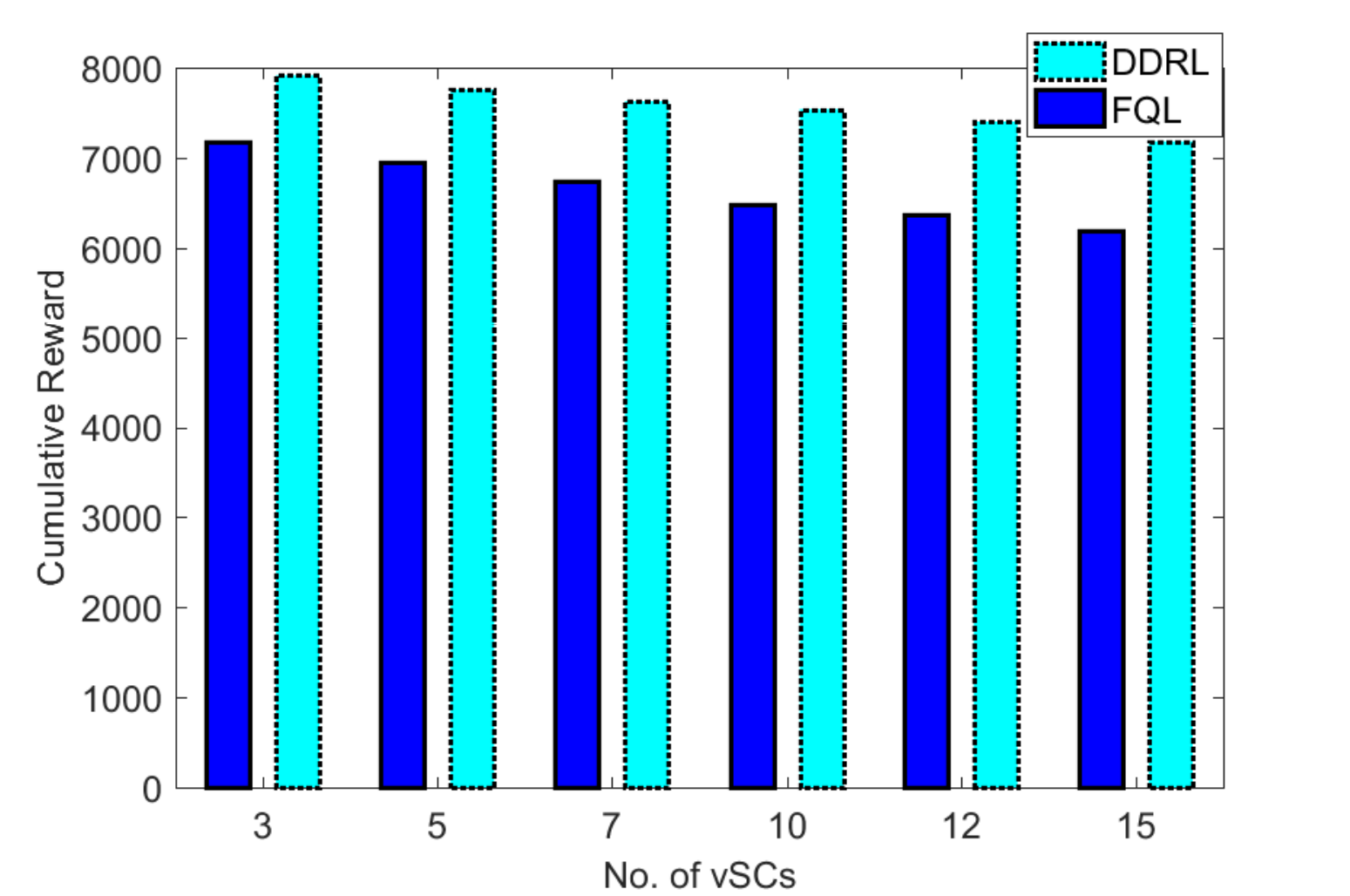}}\hspace{-0.0em}
 \vspace*{-0.25cm}
\caption{Maximum cumulative reward: (a)~Residential, (b)~Office profiles.}
\label{fig:cr_comp}
\end{figure}
The network grid energy consumption in one year of operation and the traffic drop rate comparison between DDRL and FQL controllers in residential and office areas for higher number of vSCs are shown in Figures \ref{fig:DDRL-FQL-resi} and \ref{fig:DDRL-FQL-offi}, respectively. The results show the better performance obtained by DDRL as compared to FQL controllers. More than $13\%$ and $5\%$ reduction in annual grid energy consumption is achieved with DDRL compared to FQL, in residential and office scenarios, respectively. Moreover, the DDRL control results in  up to $2.6\%$ and $1.3\%$ less traffic drop rate than FQL in office and residential scenarios, respectively. The better performance by DDRL compared to FQL is also evident from the cumulative reward achieved by the correspondent controllers in the same simulation scenarios. The maximum cumulative rewards obtained by DDRL and FQL in residential and office area traffic profiles are shown in Figure \ref{fig:cr_comp}. It shows the relatively higher cumulative reward gained by DDRL, which translates to better network performance, i.e., lower grid energy consumption and system drop rate, as justified by Figures \ref{fig:DDRL-FQL-resi} and \ref{fig:DDRL-FQL-offi}. Moreover, the gap in cumulative reward is increasing with the number of vSCs, which implies that DDRL is able to reach better coordination among the agents

\begin{table*}[h]
  \centering
\captionsetup{justification=centering}
\caption{Policy validation results.\\
{\small (R-T: Residential Training, O-T: Office Training, R-V: Residential Validation, O-V:  Office Validation)}}
\scalebox{1}{
\begin{tabular}{|c|c|c|c|c|c|c|c|c|}
\hline
\multirow{2}{*}{\centering No. of vSCs} & \multicolumn{4}{|p{4cm}|}{\centering  Grid energy consumption (KWh)} & 
    \multicolumn{4}{|p{4cm}|}{\centering Average drop rate (\%)} \\
\cline{2-9}
 & R - T & O - T & R - V & O - V & R - T & O - T & R - V & O - V\\
\hline
3 &  6874  & 6857 & 6909 &6891 & 0.00 & 0.00 & 0.00& 0.00\\

\hline
5 & 7331 & 7191 & 7412& 7287 & 0.00 &0.00 &0.00 &0.00 \\

\hline
7& 8058 & 7617 &8161  &7677 & 0.03 & 0.00 &0.10&0.10\\

\hline
10 & 8736 &8591& 8850& 8422& 1.04 &0.10 &1.61 &0.90 \\

\hline
12 & 9404 & 8775 &9508  & 8694&2.42 & 0.23 &2.38 &0.93 \\

\hline
15 & 10760 & 9116 & 10938& 9162& 2.63 & 0.61& 2.71 & 1.03\\

\hline
\end{tabular}}
\label{tab:policy-validation}
\end{table*}

\subsection{Policy Validation}
Here, we evaluate the behavior of the system in real deployment scenario after an offline training. In detail, we will validate the proposed DDRL based controllers using a new environment. \dan{Compared to the environment used for training, the new validation environment is characterized by different instantiation of both energy arrival and traffic demand. The traffic profile used for validation remains the same as the one used for training, i.e., trained in residential and validated in residential, but the particular energy harvesting and traffic generation processes used in the validation phase are different from those used in training. In particular, we use the pre-trained model with an exploration rate of  $5\%$. }

The validation of the policies along with the training environment policy evaluation for $3$, $5$, $7$, $10$, $12$ and $15$ vSCs for a year of operation are shown in Table \ref{tab:policy-validation}. The results show that DDRL agents are able to adapt their behaviors in the new environment. This is confirmed by both grid energy and average drop rate performances that are very close to the corresponding policy evaluation results. On average, only $0.5\%$ to $1.3\%$ variation is observed in the annual grid energy consumption results of policy evaluation and validation, in both residential and office profiles with no relevant changes in the traffic drop rate of the new environment. \dan{These results indicate that using an offline trained model with continuous exploration in the new environment is a viable approach for deployment. We set the exploration rate in the new environment to a small but non-zero value to allow the agents to adapt their policies in states that they have not encountered during training. We note that, since the agents are already trained, the exploration during the validation phase can be significantly smaller compared to that in the training phase, whose initial value is set to $90\%$}

\subsection{Energy Savings and Cost Analysis}
\label{sec:cost}
In Table \ref{tab:cost}, we compare the RL based controllers, i.e., FQL and DDRL, with a scenario in which all vSCs and MBS are supplied by grid power, referred as Grid-Connected (G-C). Both CAPital EXpenditures (CAPEX) and OPerational EXpenditures (OPEX) and the costs of operation for $5$ and $10$ years duration are estimated. We consider a cost of $1.17  \$/W$ for solar panels including installation, and $131  \$/KWh$ for energy storage costs \cite{nrelcost}. The energy purchasing price from the grid is set to $0.21  \$/KWh$ \cite{energyprices}.

As it is shown in Table \ref{tab:cost}, both FQL and DDRL controllers provide significant energy savings, reaching up to $54\%$ and $61\%$ for FQL and DDRL, respectively, compared to the G-C solution. In terms of cost savings, FQL provides $11\%$ and $32\%$ cost reduction during $5$ and $10$ years of operation, respectively. For DDRL controller, the cost savings rise to $17\%$ and $39\%$ during  $5$ and $10$ years of operation, respectively. Moreover, DDRL provides more energy and cost reduction than FQL controllers. These results are encouraging as they show that powering mobile networks with renewable energy sources with an intelligent control are not only environmental friendly, but also cost effective solutions. 

\begin{table*}[h]
\caption{Energy savings and costs.}
\label{tab:cost}
\centering
\begin{tabular}{lll|l|llll}
\multirow{2}{*}{\bf No. of vSCs} & \bf Algorithm &  & \multicolumn{1}{c|}{\bf Energy (kW)}        & \multicolumn{4}{c}{\bf Costs (\$)}               \\
                       &  &                         & \bf consumption [1yr] & \bf CAPEX & \bf OPEX [1yr]& \bf Cost [5yrs] & \bf Cost [10yrs]  \\ \hline \hline 
\multirow{3}{*}{3}	& G-C        &         &  11334     &   0             &   2380  &     11900     &   23801        \\ 
				& FQL 	     &         &  7136     &   2541            &   1498  &     10033     &   17526        \\ 
				& DDRL    	     &         &  6814   &    2541           &   1430  &     9695     &   16850            \\ 
\cmidrule(l){1-8} 
\multirow{3}{*}{5}	& G-C          &         &  14051      &   0             &   2950   &     14753       &   29507           \\ 
				& FQL 	     &         &  8199      &    4235           & 1721   &     12843    &   21452         \\ 
				& DDRL    	     &         & 7331     &   4235            &   1539   &     11932      &   19630            \\ 
\cmidrule(l){1-8} 
\multirow{3}{*}{7}	&  G-C         &         &  16769      &   0             &   3521   &     17607       &   35214           \\ 
				& FQL 	     &         &  8647      &   5929            &   1815   &    15008     &   24087         \\ 
				& DDRL    	     &         &  8057     &    5929           &   1691   &     14388      &   22848            \\ 
\cmidrule(l){1-8} 
\multirow{3}{*}{10}	& G-C  &         &  20845      &   0             &  4377    &    21887       &   43774          \\ 
				& FQL 	     &         &  9910      &  8470             &   2081   &     18875     &   29281         \\ 
				& DDRL    	     &         &   8736    &     8470          &   1834   &     17642      &   26815            \\ 
\cmidrule(l){1-8} 
\multirow{3}{*}{12}	& G-C&         &  23562      &   0             &     4948   &     24740       &   49480           \\ 
				& FQL 	     &         &  10844      &   10164            &   2277   &     21550     &   32936         \\ 
				& DDRL    	     &         &    9404     &     10164          &   1974   &    20038      &   29912            \\ 
\cmidrule(l){1-8} 	
\multirow{3}{*}{15}	& G-C&         &  27638      &   0             &     5803   &     29019       &   58039           \\ 
				& FQL 	     &         &  12510      &   12705            &   2627   &     25840     &   38976         \\ 
				& DDRL    	     &         &    10760     &     12705         &   2259   &    24003      &   35301            \\ 
\cmidrule(l){1-8} 	  		
\end{tabular}
\end{table*}

\section{Conclusions}
\label{sec:conclusions} 
We have proposed adaptive functional split of BB processes at vSCs powered by EH and equipped with rechargeable batteries, which can be opportunistically executed at a grid-connected edge server, co-located with the MBS. We have formulated the corresponding joint grid energy and dropped traffic minimization problem, and proposed a multi-agent DRL solution. Coordinated learning among multiple agents is enabled via the exchange of the agents' battery state information. We have evaluated the network performance, in terms of the grid energy consumption and traffic drop rate for the proposed DDRL controller, and compared the results with an offline optimization bound and a tabular MRL based controller. 
The results have confirmed that limited coordination among the agents via the exchange of battery states achieve cumulative rewards closer to the offline bounds, while requiring limited computational complexity. Extensive numerical results using traffic and EH data have confirmed that the proposed DDRL strategy ensures higher network performance, better adaptation to a changing environment, and higher cost savings with respect to the benchmark scheme. 

\dan{The work presented here can be extended in many ways. First of all, the energy saving results obtained are a good starting point to extend the solution to a scenario of very dense vSCs. However, the DDRL solution proposed here can face convergence problems in such scenarios due to non-stationarity. Moreover, the optimization can be multi-objective, e.g., minimizing energy and latency while serving the traffic demand. Hence, it may be beneficial to combine DRL with other approaches, e.g., hierarchical RL, policy based RL, or multi-objective RL, to ensure a more robust and stable controller with less sensitivity to hyperparameters. As compared to centralized approach, the DDRL solution scales better with less computational complexity (by avoiding an exponential growth in action and state spaces). However, the DDRL solution has its own limitations as the state spaces, and hence, the solutions, depend on the number of vSCs/agents. As part of our future work, we will investigate alternative solutions that are independent of the number of active agents which allows adding or removing agents in the environment without the need to re-train all the other agents.}




\section*{Acknowledgement}
This work has received funding from the European Union's Horizon 2020 research and innovation programme under the Marie \mbox{Sklodowska-Curie} grant agreement No 675891 \mbox{(SCAVENGE)} and by Spanish MINECO grant TEC2017-88373-R (5G-REFINE).

\bibliographystyle{IEEEtran}

\bibliography{ref}

\begin{thebibliography}{10}
\providecommand{\url}[1]{#1}
\csname url@samestyle\endcsname
\providecommand{\newblock}{\relax}
\providecommand{\bibinfo}[2]{#2}
\providecommand{\BIBentrySTDinterwordspacing}{\spaceskip=0pt\relax}
\providecommand{\BIBentryALTinterwordstretchfactor}{4}
\providecommand{\BIBentryALTinterwordspacing}{\spaceskip=\fontdimen2\font plus
\BIBentryALTinterwordstretchfactor\fontdimen3\font minus
  \fontdimen4\font\relax}
\providecommand{\BIBforeignlanguage}[2]{{%
\expandafter\ifx\csname l@#1\endcsname\relax
\typeout{** WARNING: IEEEtran.bst: No hyphenation pattern has been}%
\typeout{** loaded for the language `#1'. Using the pattern for}%
\typeout{** the default language instead.}%
\else
\language=\csname l@#1\endcsname
\fi
#2}}
\providecommand{\BIBdecl}{\relax}
\BIBdecl

\bibitem{cisco}
\BIBentryALTinterwordspacing
Cisco, ``Cisco visual networking index: Global mobile data traffic forecast
  update, 2017--2022,'' \emph{white paper}, 2017. [Online]. Available:
  \url{https://www.cisco.com/c/en/us/solutions/collateral/service-provider/visual-networking-index-vni/white-paper-c11-738429.pdf}
\BIBentrySTDinterwordspacing

\bibitem{footprint}
A.~Fehske, G.~Fettweis, J.~Malmodin, and G.~Biczok, ``The global footprint of
  mobile communications: The ecological and economic perspective,'' \emph{IEEE
  communications magazine}, vol.~49, no.~8, pp. 55--62, 2011.

\bibitem{ehcommag}
D.~{Gunduz}, K.~{Stamatiou}, N.~{Michelusi}, and M.~{Zorzi}, ``Designing
  intelligent energy harvesting communication systems,'' \emph{IEEE
  Communications Magazine}, vol.~52, no.~1, pp. 210--216, January 2014.

\bibitem{mec}
Y.~C. Hu, M.~Patel, D.~Sabella, N.~Sprecher, and V.~Young, ``Mobile edge
  computing a key technology towards 5g,'' \emph{ETSI white paper}, vol.~11,
  no.~11, pp. 1--16, 2015.

\bibitem{scforum}
``Virtualization for small cells: overview,'' \emph{Small cell forum}, 2015.

\bibitem{sdn-nfv}
H.~Hawilo, A.~Shami, M.~Mirahmadi, and R.~Asal, ``Nfv: state of the art,
  challenges, and implementation in next generation mobile networks (vepc),''
  \emph{IEEE Network}, vol.~28, no.~6, pp. 18--26, Nov 2014.

\bibitem{piro2013}
G.~Piro, M.~Miozzo, G.~Forte, N.~Baldo, L.~A. Grieco, G.~Boggia, and P.~Dini,
  ``Hetnets powered by renewable energy sources: Sustainable next-generation
  cellular networks,'' \emph{IEEE Internet Computing}, vol.~17, no.~1, pp.
  32--39, 2013.

\bibitem{earth-D23}
G.~Auer, O.~Blume, V.~Giannini, I.~Godor, M.~Imran, Y.~Jading, E.~Katranaras,
  M.~Olsson, D.~Sabella, P.~Skillermark \emph{et~al.}, ``Earth deliverable d2.
  3: Energy efficiency analysis of the reference systems, areas of improvements
  and target breakdown,'' \emph{Project Deliverable D}, vol.~2, 2013.

\bibitem{fss-vtc}
D.~A. {Temesgene}, N.~{Piovesan}, M.~{Miozzo}, and P.~{Dini}, ``Optimal
  placement of baseband functions for energy harvesting virtual small cells,''
  in \emph{2018 IEEE 88th Vehicular Technology Conference (VTC-Fall)}, Aug
  2018, pp. 1--6.

\bibitem{marl-survey}
L.~{Busoniu}, R.~{Babuska}, and B.~D. {Schutter}, ``A comprehensive survey of
  multiagent reinforcement learning,'' \emph{IEEE Transactions on Systems, Man,
  and Cybernetics, Part C (Applications and Reviews)}, vol.~38, no.~2, pp.
  156--172, March 2008.

\bibitem{fss-comcom}
D.~A. Temesgene, M.~Miozzo, and P.~Dini, ``Dynamic control of functional splits
  for energy harvesting virtual small cells: a distributed reinforcement
  learning approach,'' \emph{Computer Communications}, vol. 148, pp. 48--61,
  2019.

\bibitem{atari}
V.~Mnih, K.~Kavukcuoglu, D.~Silver, A.~A. Rusu, J.~Veness, M.~G. Bellemare,
  A.~Graves, M.~Riedmiller, A.~K. Fidjeland, G.~Ostrovski \emph{et~al.},
  ``Human-level control through deep reinforcement learning,'' \emph{Nature},
  vol. 518, no. 7540, p. 529, 2015.

\bibitem{piovesan2017}
N.~Piovesan and P.~Dini, ``Optimal direct load control of renewable powered
  small cells: A shortest path approach,'' \emph{Internet Technology Letters},
  2017.

\bibitem{gong2014}
J.~Gong, J.~S. Thompson, S.~Zhou, and Z.~Niu, ``Base station sleeping and
  resource allocation in renewable energy powered cellular networks,''
  \emph{IEEE Transactions on Communications}, vol.~62, no.~11, pp. 3801--3813,
  2014.

\bibitem{zhou2013}
S.~Zhou, J.~Gong, and Z.~Niu, ``Sleep control for base stations powered by
  heterogeneous energy sources,'' in \emph{2013 International Conference on ICT
  Convergence (ICTC)}.\hskip 1em plus 0.5em minus 0.4em\relax IEEE, 2013, pp.
  666--670.

\bibitem{lee2017}
G.~Lee, W.~Saad, M.~Bennis, A.~Mehbodniya, and F.~Adachi, ``Online ski rental
  for {ON/OFF} scheduling of energy harvesting base stations,'' \emph{IEEE
  Transactions on Wireless Communications}, vol.~16, no.~5, pp. 2976--2990,
  2017.

\bibitem{ehql}
P.~{Blasco}, D.~{Gunduz}, and M.~{Dohler}, ``A learning theoretic approach to
  energy harvesting communication system optimization,'' \emph{IEEE
  Transactions on Wireless Communications}, vol.~12, no.~4, pp. 1872--1882,
  April 2013.

\bibitem{miozzo2015}
M.~Miozzo, L.~Giupponi, M.~Rossi, and P.~Dini, ``Distributed q-learning for
  energy harvesting heterogeneous networks,'' in \emph{2015 IEEE International
  Conference on Communication Workshop (ICCW)}.\hskip 1em plus 0.5em minus
  0.4em\relax IEEE, 2015, pp. 2006--2011.

\bibitem{ehmab}
P.~{Blasco} and D.~{Gunduz}, ``Multi-access communications with energy
  harvesting: A multi-armed bandit model and the optimality of the myopic
  policy,'' \emph{IEEE Journal on Selected Areas in Communications}, vol.~33,
  no.~3, pp. 585--597, March 2015.

\bibitem{ameur2016}
H.~Ameur, M.~Esseghir, and L.~Khoukhi, ``Fully distributed approach for energy
  saving in heterogeneous networks,'' in \emph{2016 8th IFIP international
  conference on New Technologies, Mobility and Security (NTMS)}.\hskip 1em plus
  0.5em minus 0.4em\relax IEEE, 2016, pp. 1--6.

\bibitem{ll-vtc}
M.~{Miozzo} and P.~{Dini}, ``Layered learning radio resource management for
  energy harvesting small base stations,'' in \emph{2018 IEEE 87th Vehicular
  Technology Conference (VTC Spring)}, June 2018, pp. 1--6.

\bibitem{LL}
M.~{Miozzo}, N.~{Piovesan}, and P.~{Dini}, ``Coordinated load control of
  renewable powered small base stations through layered learning,'' \emph{IEEE
  Transactions on Green Communications and Networking}, pp. 1--1, 2019.

\bibitem{drag}
Y.~Junhong and Y.~J. Zhang, ``Drag: Deep reinforcement learning based base
  station activation in heterogeneous networks,'' \emph{IEEE Transactions on
  Mobile Computing}, 2019.

\bibitem{mendil}
M.~{Mendil}, A.~{De Domenico}, V.~{Heiries}, R.~{Caire}, and N.~{Hadjsaid},
  ``Battery-aware optimization of green small cells: Sizing and energy
  management,'' \emph{IEEE Transactions on Green Communications and
  Networking}, vol.~2, no.~3, pp. 635--651, Sep. 2018.

\bibitem{xu2016}
J.~Xu and S.~Ren, ``Online learning for offloading and autoscaling in
  renewable-powered mobile edge computing,'' in \emph{Global Communications
  Conference (GLOBECOM), 2016 IEEE}.\hskip 1em plus 0.5em minus 0.4em\relax
  IEEE, 2016, pp. 1--6.

\bibitem{dynamic-pimrc}
D.~A. {Temesgene}, M.~{Miozzo}, and P.~{Dini}, ``Dynamic functional split
  selection in energy harvesting virtual small cells using temporal difference
  learning,'' in \emph{2018 IEEE 29th Annual International Symposium on
  Personal, Indoor and Mobile Radio Communications (PIMRC)}, Sep. 2018, pp.
  1813--1819.

\bibitem{wangflexible}
L.~Wang and S.~Zhou, ``Flexible functional split and power control for energy
  harvesting cloud radio access networks,'' \emph{IEEE Transactions on Wireless
  Communications}, 2019.

\bibitem{ngmn}
\BIBentryALTinterwordspacing
N.~Alliance, ``Ngmn kpis and deployment scenarios for consideration for
  imt2020,'' \emph{Final Deliverable}, 2016. [Online]. Available:
  \url{https://www.ngmn.org}
\BIBentrySTDinterwordspacing

\bibitem{3gpp-38801}
3GPP, ``{TS 38.801.; ; {E-U}; Study on new radio access technology: Radio
  access architecture and interfaces v14},'' 2017.

\bibitem{Lu2013}
L.~Lu, X.~Han, J.~Li, J.~Hua, and M.~Ouyang, ``A review on the key issues for
  lithium-ion battery management in electric vehicles,'' \emph{Journal of power
  sources}, vol. 226, pp. 272--288, 2013.

\bibitem{desset2012}
C.~Desset, B.~Debaillie, V.~Giannini, A.~Fehske, G.~Auer, H.~Holtkamp,
  W.~Wajda, D.~Sabella, F.~Richter, M.~J. Gonzalez \emph{et~al.}, ``Flexible
  power modeling of lte base stations,'' in \emph{2012 IEEE wireless
  communications and networking conference (WCNC)}.\hskip 1em plus 0.5em minus
  0.4em\relax IEEE, 2012, pp. 2858--2862.

\bibitem{Miozzo2014}
M.~Miozzo, D.~Zordan, P.~Dini, and M.~Rossi, ``Solarstat: Modeling photovoltaic
  sources through stochastic markov processes,'' in \emph{2014 IEEE
  International Energy Conference (ENERGYCON)}.\hskip 1em plus 0.5em minus
  0.4em\relax IEEE, 2014, pp. 688--695.

\bibitem{Xu2017understanding}
F.~Xu, Y.~Li, H.~Wang, P.~Zhang, and D.~Jin, ``Understanding mobile traffic
  patterns of large scale cellular towers in urban environment,''
  \emph{IEEE/ACM transactions on networking (TON)}, vol.~25, no.~2, pp.
  1147--1161, 2017.

\bibitem{mobiletraffic}
H.~D. {Trinh}, L.~{Giupponi}, and P.~{Dini}, ``Mobile traffic prediction from
  raw data using lstm networks,'' in \emph{2018 IEEE 29th Annual International
  Symposium on Personal, Indoor and Mobile Radio Communications (PIMRC)}, Sep.
  2018, pp. 1827--1832.

\bibitem{suttonRL}
R.~S. Sutton and A.~G. Barto, \emph{Reinforcement learning: An
  introduction}.\hskip 1em plus 0.5em minus 0.4em\relax MIT press Cambridge,
  1998, vol.~1, no.~1.

\bibitem{cyclic}
A.~Adams and P.~Vamplew, ``Encoding and decoding cyclic data,'' \emph{The South
  Pacific Journal of Natural and Applied Sciences}, 1998.

\bibitem{auer2010}
G.~Auer, O.~Blume, V.~Giannini, I.~Godor, M.~Imran, Y.~Jading, E.~Katranaras,
  M.~Olsson, D.~Sabella, P.~Skillermark \emph{et~al.}, ``D2. 3: Energy
  efficiency analysis of the reference systems, areas of improvements and
  target breakdown,'' \emph{EARTH}, vol.~20, no.~10, 2010.

\bibitem{lces}
N.~{Piovesan}, D.~A. {Temesgene}, M.~{Miozzo}, and P.~{Dini}, ``Joint load
  control and energy sharing for autonomous operation of 5g mobile networks in
  micro-grids,'' \emph{IEEE Access}, vol.~7, pp. 31\,140--31\,150, 2019.

\bibitem{DaSilva2018}
A.~P. {Couto da Silva}, D.~{Renga}, M.~{Meo}, and M.~{Ajmone Marsan}, ``The
  impact of quantization on the design of solar power systems for cellular base
  stations,'' \emph{IEEE Transactions on Green Communications and Networking},
  vol.~2, no.~1, pp. 260--274, March 2018.

\bibitem{nrelcost}
R.~Fu, D.~J. Feldman, and R.~M. Margolis, ``Us solar photovoltaic system cost
  benchmark: Q1 2018,'' National Renewable Energy Lab. (NREL), United States,
  Tech. Rep., 2018.

\bibitem{energyprices}
\BIBentryALTinterwordspacing
\emph{Average Energy Prices}, 2019 (accessed Oct., 2019). [Online]. Available:
  \url{http://bit.ly/averagepricesLA}
\BIBentrySTDinterwordspacing

\end{thebibliography}

\enlargethispage{0in}
\begin{IEEEbiography}
[{\includegraphics[width=1in,height=1.25in,clip,keepaspectratio]{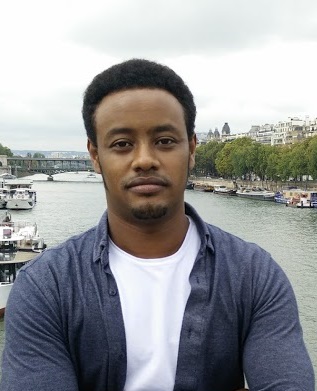}}]{Dagnachew Azene Temesgene}
received B.Sc. degree in Electrical and Computer Engineering from Bahir Dar University, Ethiopia, European joint M.Sc. degree in Pervasive Computing and Communications in 2016 and Ph.D. in Network Engineering from Universitat Politècnica de Catalunya - BarcelonaTech (UPC) in 2020. While pursuing his Ph.D, he worked as a research assistant at Centre Tecnològic de Telecomunicacions de Catalunya (CTTC) in the framework of the EU H2020 Marie Skłodowska-Curie actions (MSCA) SCAVENGE project. He is currently a researcher at Ericsson in Stockholm, Sweden. His research interests include mobile networks, energy harvesting communications and machine learning for network control and management.
\end{IEEEbiography}
\vskip -7pt plus -1fil
\begin{IEEEbiography}
[{\includegraphics[width=1in,height=1.25in,clip,keepaspectratio]{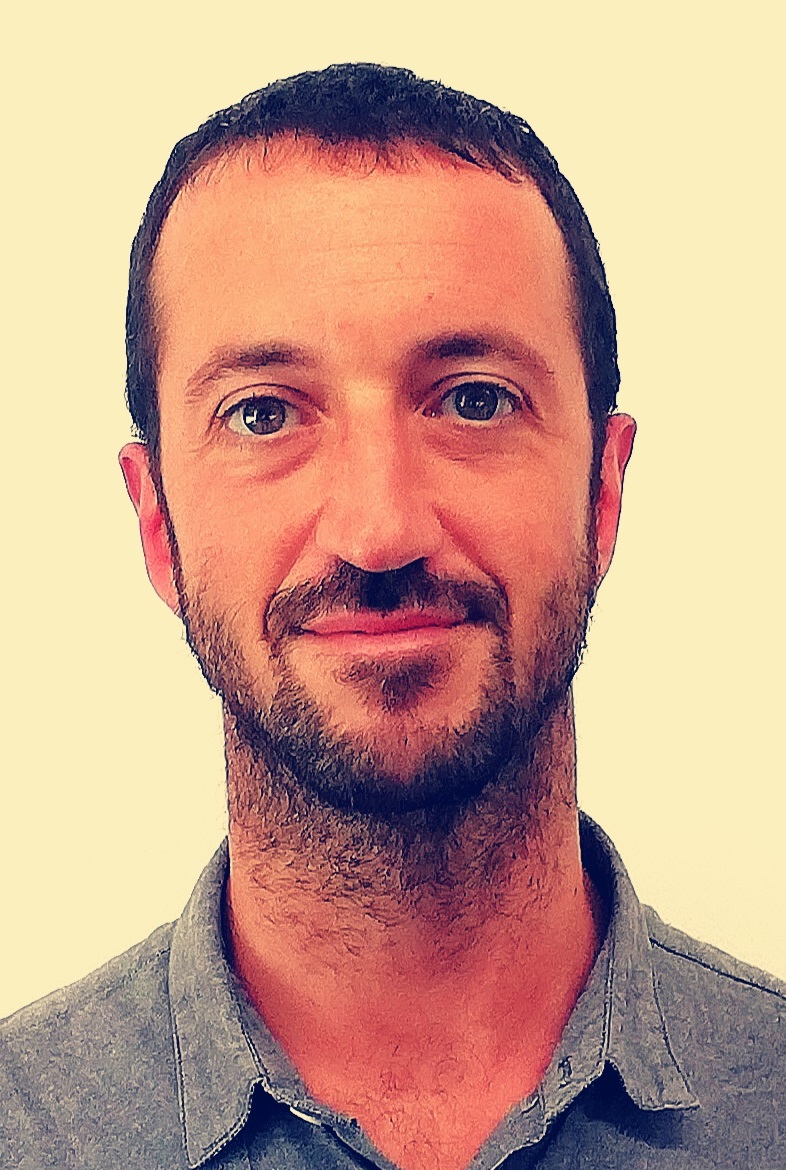}}]{Marco Miozzo}
received his M.Sc. degree in Telecommunication Engineering from the University of Ferrara (Italy) in 2005 and the Ph.D. from the Technical University of Catalonia (UPC) in 2018. 
In June 2008 he joined the Centre Tecnològic de Telecomunicacions de Catalunya (CTTC). In CTTC he has been involved in the development of an LTE module for the network simulator 3 (ns-3) in the framework of the LENA project. Currently he is collaborating with the EU founded H2020  SCAVENGE (MSCA ETN). He participated in several R\&D projects and he has been involved in 5G-Crosshaul, Flex5Gware and SANSA projects, working on environmental sustainable mobile networks with energy harvesting capabilities through learning techniques. His main research interests are: sustainable mobile networks, green wireless networking, energy harvesting, multi-agent systems, machine learning, green AI, energy ethical consumerism, transparent and explainable AI.
\end{IEEEbiography}
\vskip 0pt plus -1fil
\begin{IEEEbiography}
[{\includegraphics[width=1in,height=1.25in,clip,keepaspectratio]{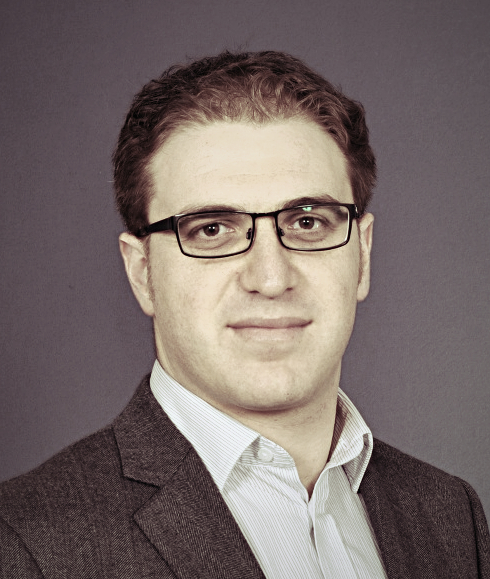}}]{Deniz G{\"u}nd{\"u}z}
[S’03-M’08-SM’13] received the M.S. and Ph.D. degrees in electrical engineering from the NYU Polytechnic School of Engineering in 2004 and 2007, respectively. He served as a Post-Doctoral Research Associate at Princeton University, as a Consulting Assistant Professor at
Stanford University, and as a Research Associate at the Centre Tecnologic de Telecommunicaciones de Catalunya (CTTC). He is currently a Professor in the Electrical and Electronic Engineering Department, Imperial College London, U.K., where he leads the Information Processing and Communications Laboratory. His research interests lie in the areas of communications and information theory, machine learning, and security and privacy in cyber-physical systems. He is a Distinguished Speaker of the IEEE Information Theory Society. He is a recipient of the IEEE Communications Society– Communication Theory Technical Committee Early Achievement Award in 2017 and a Starting Grant of the European Research Council in 2016. He has co-authored papers that received the Best Paper Award at the 2016 IEEE WCNC and 2019 IEEE GlobalSIP, and best student paper awards at 2007 IEEE ISIT and 2018 IEEE WCNC. He is an Editor of the IEEE TRANSACTIONS ON WIRELESS COMMUNICATIONS and an Area Editor for the IEEE TRANSACTIONS ON COMMUNICATIONS.
\end{IEEEbiography}
\vskip 0pt plus -1fil
\begin{IEEEbiography}
[{\includegraphics[width=1in,height=1.25in,clip,keepaspectratio]{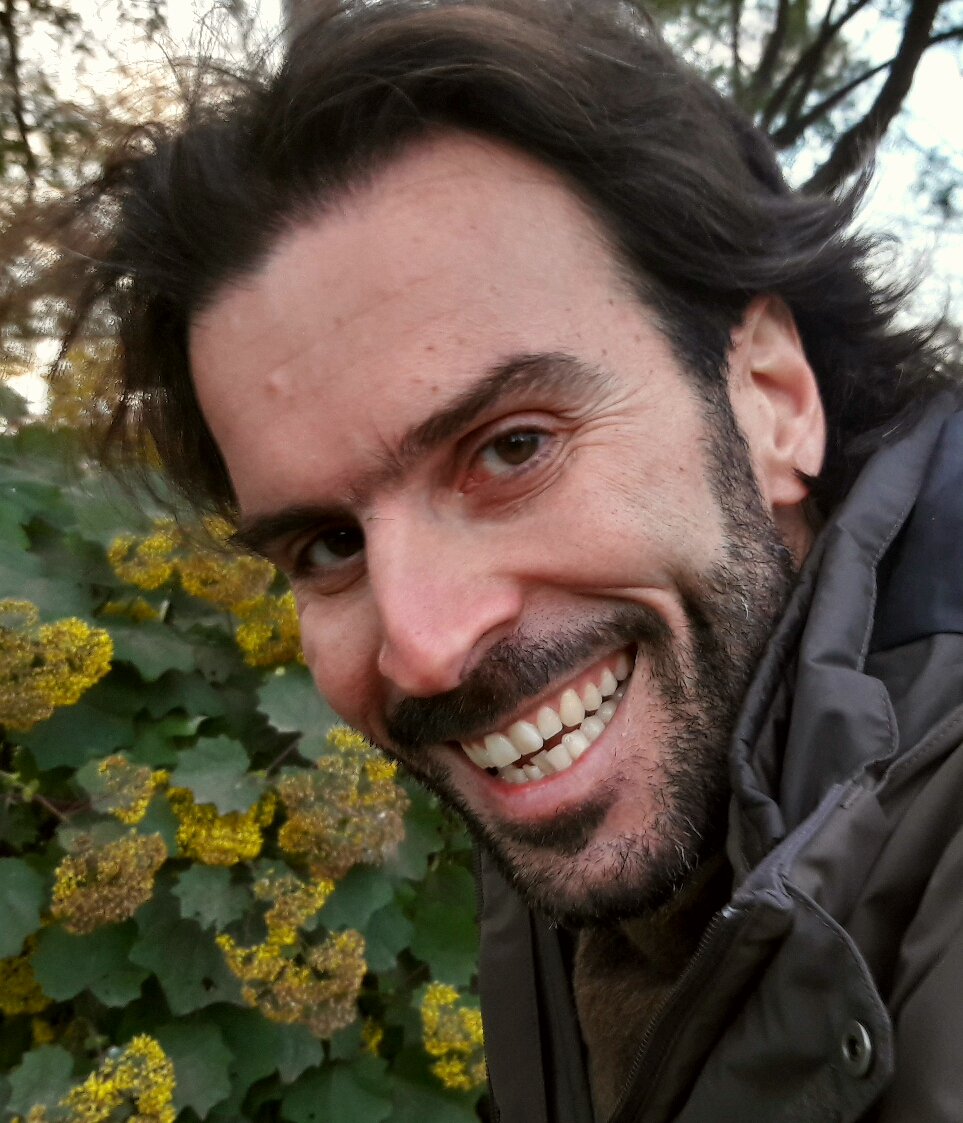}}]{Paolo Dini}
received MSc and PhD from Universit{\`a} di Roma La Sapienza, in 2001 and 2005, respectively. He served as a Post-Doc Researcher at the Research Centre on Software Technology (RCOST) - Universit{\`a} del Sannio, and contracted Professor at Universit{\`a} di Roma La Sapienza in 2005. He is now with the Centre Tecnologic de Telecomunicacions de Catalunya (CTTC) as a Senior Researcher. He received two awards from the Cisco Silicon Valley Foundation for his research on heterogeneous mobile networks in 2008 and 2011, respectively. He has been involved in over 25 research projects related to network management, optimization and energy efficiency. He is currently the Coordinator of the EU H2020 MSCA SCAVENGE European Training Network on sustainable mobile networks with energy harvesting capabilities. His research interests include sustainable networking and computing, distributed optimization and optimal control, machine learning and data analytics.
\end{IEEEbiography}
\end{document}